\pdfoutput=1

\documentclass[11pt,twoside,a4paper,cmspaper,final,collab]{cms-tdr}
\def\svnVersion{318646}\def\cmsCernNo{2013-037}\def\cmsCernDate{\today}\def\cmsMessage{Published in the Journal of High Energy Physics as \href{http://dx.doi.org/10.1007/JHEP01(2016)079}{\doi{10.1007/JHEP01(2016)079.}}}

\begin{document}\cmsNoteHeader{HIG-14-019}

\hyphenation{had-ron-i-za-tion}
\hyphenation{cal-or-i-me-ter}
\hyphenation{de-vices}
\RCS$Revision: 313573 $
\RCS$HeadURL: svn+ssh://svn.cern.ch/reps/tdr2/papers/HIG-14-019/trunk/HIG-14-019.tex $
\RCS$Id: HIG-14-019.tex 313573 2015-12-04 22:05:03Z rasp $
\newlength\cmsFigWidth
\ifthenelse{\boolean{cms@external}}{\setlength\cmsFigWidth{0.85\columnwidth}}{\setlength\cmsFigWidth{0.4\textwidth}}
\ifthenelse{\boolean{cms@external}}{\providecommand{\cmsLeft}{top}}{\providecommand{\cmsLeft}{left}}
\ifthenelse{\boolean{cms@external}}{\providecommand{\cmsRight}{bottom}}{\providecommand{\cmsRight}{right}}

\newcommand{\PaO}{\ensuremath{\mathrm{a}_1}}
\newcommand{\PaTw}{\ensuremath{\mathrm{a}_2}}
\newcommand{\PhO}{\ensuremath{\mathrm{h}_1}}
\newcommand{\PhTw}{\ensuremath{\mathrm{h}_2}}
\newcommand{\PhTh}{\ensuremath{\mathrm{h}_3}}
\newcommand{\PhQ}{\ensuremath{\mathrm{h}^{\pm}}}
\newcommand{\sigmaBR}{\ensuremath{(\sigma \mathcal{B})_\text{sig}}}

\cmsNoteHeader{HIG-14-019}
\title{Search for a very light NMSSM Higgs boson produced in decays of the 125 GeV scalar boson 
and decaying into \texorpdfstring{$\Pgt $~leptons}{tau leptons} in pp collisions at \texorpdfstring{$\sqrt{s}=8\TeV$}{sqrt(s) = 8 TeV}}

\date{\today}

\abstract{
A search for a very light Higgs boson decaying into a pair of $\Pgt$ leptons is presented within the framework of the next-to-minimal supersymmetric standard model. This search is based on a data set corresponding to an  integrated luminosity of 19.7\fbinv of proton-proton collisions collected by the CMS experiment at a centre-of-mass energy of 8\TeV. The signal is defined by the production of either of the two lightest scalars, $\PhO$ or $\PhTw$, via gluon-gluon fusion and subsequent decay into a pair of the lightest Higgs bosons, $\PaO$ or $\PhO$. The $\PhO$ or $\PhTw$ boson is identified with the observed state at a mass of 125\GeV.  The analysis searches for decays of the $\PaO\ (\PhO)$ states into pairs of $\Pgt$ leptons and covers a mass range for the $\PaO\ (\PhO)$ boson of 4 to 8\GeV. The search reveals no significant excess in data above standard model background expectations, and an upper limit is set on the signal production cross section times branching fraction as a function of the $\PaO\ (\PhO)$ boson mass.  The 95\% confidence level limit ranges from 4.5\unit{pb} at $m_{\PaO}\ (m_{\PhO})=8\GeV$ to 10.3\unit{pb} at $m_{\PaO}\ (m_{\PhO})=5\GeV$.      
}

\hypersetup{%
pdfauthor={CMS Collaboration},%
pdftitle={Search for a very light NMSSM Higgs boson produced in decays of the 125 GeV scalar boson and decaying into tau leptons in pp collisions at sqrt(s) = 8 TeV},%
pdfsubject={CMS},%
pdfkeywords={CMS, physics, Higgs boson, NMSSM}}

\maketitle

\section{Introduction}
\label{sec:intro}

The recently discovered particle with mass close to 125\GeV~\cite{Aad:2012tfa,Chatrchyan:2012ufa,Chatrchyan:DiscoveryLongPaper}
has been shown to have properties that are consistent with those of a
standard model (SM) Higgs 
boson~\cite{CMS_Higgs_properties,CMS_Hgammagamma,CMS_HZZ,CMS_HWW,CMS_Htautau,CMS_Hfermions,ATLAS_Hgammagamma,ATLAS_HZZ,ATLAS_HWW,ATLAS_Htautau,ATLAS_mass}.
Supersymmetric (SUSY) extensions of the SM~\cite{SUSY:1,SUSY:2} also predict a particle with such properties and resolve some 
problems of the SM~\cite{Ellis:2002wba}. The minimal supersymmetric standard model (MSSM)~\cite{MSSM:1,MSSM:2} postulates the existence of
two Higgs doublets, resulting in five physical states: two CP-even, one CP-odd, and two charged Higgs bosons.
This version of SUSY has been extensively tested using data collected by the ATLAS and CMS
experiments at the CERN LHC. However, nonminimal SUSY extensions have received far less
attention. One example is the next-to-MSSM (NMSSM), which extends the
MSSM by an additional singlet superfield, interacting only with itself
and the two Higgs doublets~\cite{MSSM:1,Kaul198236,Barbieri1982343,Nilles1983346,Frere198311,Derendinger1984307,Ellwanger:2009dp,Maniatis:2009re}. 
This scenario has all the desirable features of SUSY, including a solution of the
hierarchy problem and gauge coupling unification. In the NMSSM, the Higgs mixing parameter $\mu$ 
is naturally generated at the electroweak scale through the vacuum expectation value of the singlet 
field, thereby solving the so-called $\mu$ problem of the MSSM~\cite{Kim:1983dt}. Furthermore, the amount
of fine tuning required in the NMSSM to obtain a CP-even 
Higgs boson with a mass of 125\GeV is significantly reduced compared to the MSSM~\cite{Casas:2004,Dermisek:2005ar,Dermisek:2007yt}.
The Higgs sector of the NMSSM is larger than
that of the MSSM. There are seven Higgs bosons: three CP-even $(\mathrm{h}_{1, 2, 3})$,
two CP-odd $(\mathrm{a}_{1, 2})$, and two charged Higgs states.  
By definition, $m_{\PhTh} > m_{\PhTw} > m_{\PhO}$ and $m_{\PaTw} > m_{\PaO}$. Over
large parts of the NMSSM parameter space, the observed boson with mass
close to 125\GeV, hereafter denoted $\PH(125)$, could be identified with
one of the two lightest scalar NMSSM Higgs bosons, \PhO\ or \PhTw.

A vast set of next-to-minimal supersymmetric models
is consistent with the SM measurements and constraints from searches for SUSY particles made with LHC,
Tevatron, SLAC and LEP data, as well as 
with the properties of the $\PH(125)$ boson measured using Run~1 LHC data~\cite{Belanger:2012tt,Belanger:2012sd,Gunion:2012he,Gunion:2012gc,King:2012is,King:2012tr}.  
These models provide possible signatures that cannot be realized in 
the MSSM given recent experimental constraints~\cite{Curtin:2014}. For example, 
the decays $\PH(125)\to \PhO\PhO$ and $\PH(125)\to \PaO\PaO$ are allowed when kinematically possible.
These decay signatures 
have been investigated in phenomenological studies
considering a variety of production modes at the 
LHC~\cite{Ellwanger:2005uu,Ellwanger:2003jt,Ellwanger:2004gz,Belyaev:2008gj,Belyaev:2010ka,Lisanti:2009uy,Almarashi:2012ri,Almarashi:2011qq}.
The analysis presented in this paper is motivated
by the NMSSM scenarios that predict a very light \PhO\ or \PaO\ state with
mass in the range $2m_{\Pgt}<m_{\PhO}~(m_{\PaO})<2m_{\PQb}$, where
$m_{\Pgt}$ is the mass of the \Pgt\ lepton and $m_{\PQb}$ is the mass of
the \PQb quark. Such a light state is potentially accessible in  
final states with four \Pgt~leptons, where $\PH(125)\to \PhO\PhO\,(\PaO\PaO) \to 4\Pgt$~\cite{Bomark:2014qua,King:2014xwa}.
In these scenarios the decay $\PH(125)\to \PaTw\PaTw$ is not kinematically allowed.

Several searches for $\PH(125)\to \phi_1\phi_1$ decays, where $\phi_1$
can be either the lightest CP-even state \PhO\ or the lightest
CP-odd state \PaO, have been performed. The analyses carried out by the 
OPAL and ALEPH Collaborations at LEP~\cite{Abbiendi:2002in, Schael:2010aw}
searched for the decay of the CP-even Higgs boson into a pair of light CP-odd Higgs bosons,
exploiting the Higgs-strahlung process, where the CP-even state is produced in 
association with a \cPZ\ boson. These searches found no evidence for a signal, and limits
were placed on the signal production cross section times branching fraction. However, searches at LEP
did not probe masses of the CP-even state above 114\GeV. 
A similar study has been performed 
by the \DZERO Collaboration at the Tevatron~\cite{Abazov:2009yi}, searching for inclusive production of the 
CP-even Higgs boson in $\Pp\Pap$ collisions followed by its decay into a pair of light CP-odd Higgs bosons. 
No signal was detected and upper limits were set
on the signal production cross section times branching fraction
in the mass ranges $3.6 < m_{\PaO} < 19\GeV$
and $89 < m_{\PH} < 200\GeV$. 
The limits set by the \DZERO analysis are a factor one to seven times higher compared to the SM production cross section 
for $\Pp\Pap\to\PH(125)+\rm{X}$.

The CMS Collaboration has recently searched for a very light CP-odd 
Higgs boson produced in decays of a heavier CP-even
state~\cite{CMS-PAS-HIG-13-010}. This study probed the mass of the
CP-odd state in the range $2m_{\Pgm} < m_{\PaO} < 2m_{\Pgt}$, where $m_{\Pgm}$ is the mass of muon. 
In this mass range the decay
$\PaO \to \Pgm\Pgm$ can be significant.  No evidence for a signal was found and
upper limits were placed on the signal production cross section 
times branching fraction.
The ATLAS Collaboration has also recently searched for $\mathrm{h}/\PH \to \PaO\PaO \to \Pgm\Pgm\Pgt\Pgt$~\cite{Aad:2015oqa},
covering the mass range $m_{\PaO} = 3.7$--50\GeV for $m_{\PH} = 125\GeV$, and $m_{\PH} = 100$--500\GeV for $m_{\PaO} = 5\GeV$.
No excess over SM backgrounds was observed, and upper limits were placed on 
$\sigma(\cPg\cPg \to \PH) \, \mathcal{B}(\PH \to \PaO\PaO) \, \mathcal{B}^{2}(\PaO \to \Pgt \Pgt)$,
under the assumption that 

\begin{displaymath}
\frac{\Gamma({\rm{a}} \to \Pgm\Pgm)}{\Gamma({\rm{a}} \to \Pgt\Pgt)}
=\frac{m^2_{\Pgm}}{m^2_{\Pgt}\sqrt{1-\big(2m_{\Pgt}/m_{\rm{a}}\big)^2}}.
\end{displaymath}

The search for the production of a pair of light bosons with their subsequent decay into four \Pgt\ leptons has
not yet been performed at the LHC and is the subject of this paper. The choice of the 4\Pgt\ channel
makes it possible to probe the signal cross section times branching fraction

\begin{displaymath}
(\sigma \mathcal{B})_\text{sig} \equiv \sigma (\cPg\cPg \to \PH(125)) \, \mathcal{B} (\PH(125) \to \phi_1\phi_1) \, \mathcal{B}^{2} (\phi_1 \to \Pgt \Pgt)
\end{displaymath}

in a model-independent way.

\section{Signal topology}
\label{Sec:Topology}
This paper describes a search for the production of the $\PH(125)$ boson,
with its decay into a pair of light NMSSM Higgs bosons $\phi_1$.
The signal can be associated with one
of three possible scenarios:

\begin{itemize}
\item {$\PH(125)$ corresponds to \PhTw\ and decays into a pair of \PhO\ states, $\PhTw \to \PhO\PhO$;}
\item {$\PH(125)$ corresponds to \PhTw\ and decays into a pair of \PaO\ states, $\PhTw \to \PaO\PaO$;}
\item {$\PH(125)$ corresponds to \PhO\ and decays into a pair of \PaO\ states, $\PhO \to \PaO\PaO$.}
\end{itemize}

\begin{figure*}[hbtp]
\begin{center}
\includegraphics[width=0.45\textwidth]{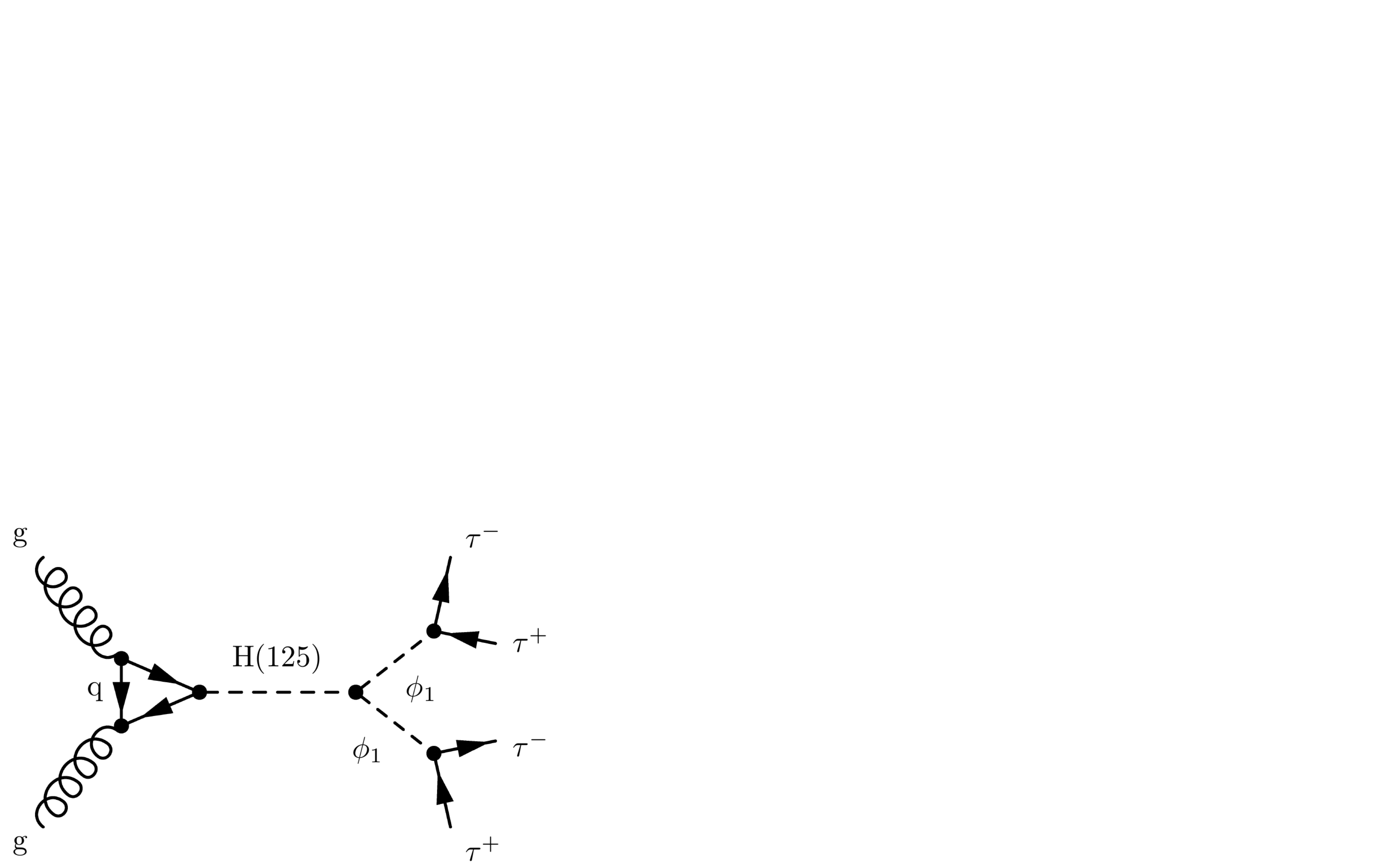}\hspace{1cm}
\includegraphics[width=0.45\textwidth]{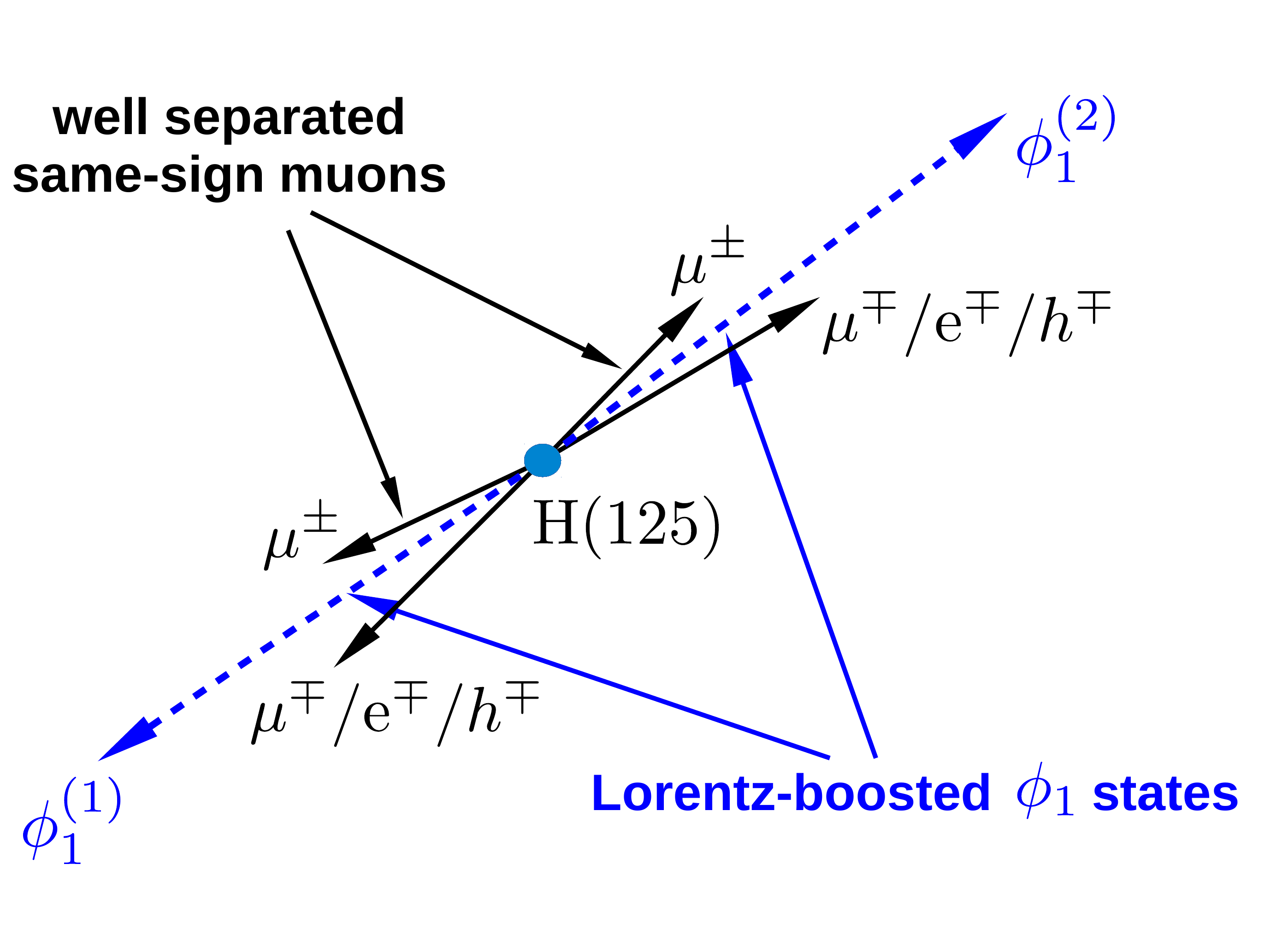}
\caption{
Left: Feynman diagram for the signal process. Right: Illustration of the signal topology. The label
``$\Pgm^\mp/\mathrm{e}^\mp/h^\mp$'' denotes a muon, electron, or charged-hadron track. 
}
\label{fig:topology}
\end{center}
\end{figure*}

The analysis is optimized for the gluon-gluon fusion
process, which is the dominant production mechanism of the $\PH(125)$~boson at the LHC.
The signal topology is illustrated in Fig.~\ref{fig:topology}.
The search is performed for very light $\phi_1$ states, 
covering a mass range of 4 to 8\GeV.
Within this mass range the $\phi_1$ boson is expected to 
decay predominantly into a pair of \Pgt\ leptons,
$\phi_1\to\Pgt\Pgt$. In the decay of each $\phi_1$, one of the \Pgt\ leptons is identified via its
muonic decay. The other \Pgt\ lepton is required to decay into a one-prong mode,
\ie a decay into one charged particle (electron, muon, or hadron)
and one or more neutral particles. We identify these decays by the presence 
of one reconstructed track with charge sign opposite to that of the closest muon. 
Neutral particles are not considered in the event selection.

Given the large difference in mass between the $\phi_1$ and the
$\PH(125)$ states ($m_{\PH(125)}\gg m_{\phi_1}$), one expects the
$\phi_1$ bosons to have large Lorentz boosts and
their decay products to be collimated. Furthermore, in the gluon-gluon fusion process the 
$\PH(125)$ state is mainly produced with relatively small transverse
momentum \pt. Thus, in the majority of $\PH(125)\to \phi_{1}\phi_{1}$
decays, the $\phi_{1}$ states would be produced nearly back-to-back in the plane transverse to the beams,
with a large separation in azimuthal angle $\phi$ between the decay products of the two $\phi_1$ bosons. 
The $\PH(125)$ state can be produced with relatively high transverse momentum if a hard
gluon is radiated from the initial-state gluons or the heavy-quark loop.
In this case, the separation between two $\phi_1$ bosons in azimuthal angle is reduced, while
the separation in pseudorapidity $\eta$ can still be large. The pseudorapidity
is defined via the polar angle $\theta$ as $\eta\equiv -\ln\left[\tan{(\theta/2)}\right]$.

The $\phi_1\rightarrow\tau\tau$ decays into final states without muons
are not considered in the analysis. These decays are mimicked by hadronic jets
with a significantly higher probability compared to final states with at least one muon
and contribute marginally to the search sensitivity.

The signal properties discussed above are used to define the search topology.
The analysis presented here searches for the signal in a sample of dimuon events with large angular separation between the muons.
The two muons are required to have the same sign. This criterion almost entirely eliminates background from the Drell--Yan process, 
gauge boson pair production, and \ttbar production. 
Each muon is accompanied by one nearby opposite-sign track.
Further details of the kinematic selection are given in Section~\ref{Sec:Selection}. 
Throughout this paper, the signal yields are normalized to the benchmark value of the signal production cross section 
times branching fraction of 5\unit{pb}.
The choice of the benchmark scenario is motivated by recent phenomenological analyses~\cite{King:2014xwa, Bomark:2014qua}.

\section{CMS detector, data, and simulated samples}
\label{sec:detector}

The central feature of the CMS apparatus is a superconducting solenoid of 6\unit{m} internal diameter, providing a field of 3.8\unit{T}. 
The innermost component of the detector is a silicon pixel and strip tracker, which is used to measure the momenta of charged particles and 
reconstruct collision vertices. The tracker, which covers the pseudorapidity range $|\eta|<2.5$, 
is surrounded by a crystal electromagnetic 
calorimeter and a brass and scintillator hadronic calorimeter, both placed inside the solenoid. These calorimeters cover
$|\eta|<3.0$. A quartz fiber Cherenkov forward hadron detector extends the calorimetric coverage to $|\eta|<5.0$. Muons are measured in the 
pseudorapidity range $\abs{\eta}< 2.4$, with detection planes made using three technologies: drift tubes, cathode strip chambers, and 
resistive-plate chambers. 
The first level of the CMS trigger system, composed of custom hardware processors, uses information from the calorimeters and
muon detectors to select the most interesting events in a fixed time interval of 4\mus. The high-level trigger processor farm further 
decreases the event rate from around 100\unit{kHz} to less than 1\unit{kHz} before data storage.
A more detailed description of the CMS detector, together with a definition of the coordinate system used and the relevant 
kinematic variables, can be found in Ref.~\cite{Chatrchyan:2008zzk}.

The data set used in this analysis was recorded in 2012
and corresponds to an integrated luminosity of 19.7\fbinv
of $\Pp\Pp$ collisions at $\sqrt{s}=8\TeV$.

The Monte Carlo (MC) event generator \PYTHIA 6.426~\cite{PYTHIA} is used to model the NMSSM Higgs boson signal produced via gluon-gluon fusion. 
The $\PH(125)$ boson \pt spectrum from \PYTHIA is reweighted to the spectrum obtained from a next-to-leading-order computation with a 
next-to-next-to-leading logarithmic accuracy using the {\sc HqT} 2.0 program~\cite{Hqt1,Hqt2}, which performs the resummation of the 
large logarithmic contributions appearing at transverse momenta much smaller than the mass of the Higgs boson. 
For optimisation studies, diboson and quantum chromodynamics (QCD) multijet backgrounds are simulated by \PYTHIA. Inclusive \cPZ, \PW, 
and \ttbar production are modelled with \MADGRAPH 5.1~\cite{MADGRAPH}. 
The \MADGRAPH generator is interfaced with \PYTHIA for parton showering and fragmentation. 
The \PYTHIA parameters that steer the simulation of
hadronisation and the underlying event
are set to the most recent \PYTHIA Z2* tune. 
This tune is derived from the Z1 tune~\cite{Z1tune}, which uses the CTEQ5L parton distribution function (PDF) set, 
whereas Z2* adopts the CTEQ6L PDF set~\cite{CTEQ6L1}. 
The \TAUOLA  package~\cite{TAUOLA} is used for \Pgt\ lepton decays in all cases.
All generated events, with the exception of a few special QCD multijet samples discussed in Section 6.2,
are processed through a detailed simulation of the CMS detector, based on \GEANTfour~\cite{Agostinelli2003250}, 
and are reconstructed employing the same algorithms as for data.

\section{Event selection}
\label{Sec:Selection}

Events are recorded using double-muon triggers with thresholds on the muon 
transverse momenta of 17\GeV for the leading muon and 8\GeV for the subleading one.
To pass the high-level trigger, the tracks of the two muons are additionally required 
to have points of closest approach to the beam axis within 2\mm of each other 
along the longitudinal direction.

In 2012, the average number of $\Pp\Pp$ interactions
per LHC bunch crossing (pileup) was about 20.
The simulated MC events are reweighted to represent the distribution of the number of 
pileup interactions per bunch crossing in data.

For each reconstructed collision vertex, the sum of the ${\pt}^2$ of all tracks
associated with the vertex is computed. The vertex for which this quantity is largest 
is assumed to correspond to the hard-scattering process, and is referred to
as the primary vertex (PV). 

The identification and reconstruction of muons is achieved by matching track segments found
in the silicon tracker with those found in the muon detectors~\cite{CMS_Muon_Reco}. Additional requirements
are applied on the number of measurements in the inner pixel and
outer silicon strip detectors, on the number of matched segments in the muon detectors, and
on the quality of the global muon track fit, quantified by $\chi^2$.

The data are further selected by requiring at least one pair of muons with the same charge.
This requirement significantly suppresses
background contributions originating from the Drell--Yan process, from decays of
\ttbar pairs, and from QCD multijet events
with muonic decays of heavy-flavour hadrons. 
The leading muon is required to have $\pt > 17\GeV$ and $\abs{\eta} < 2.1$.
The subleading muon is required to have $\pt > 10\GeV$ and $\abs{\eta} < 2.1$.
To reject QCD multijet events with muonic decays of hadrons containing charm or bottom
quarks, selections are applied on the impact parameters of the muon tracks. 
The impact parameter in the transverse plane is required to be smaller than 300\micron with respect to the PV.
The longitudinal impact parameter is required to be smaller than 1\mm with respect to the PV.
The two selected same-sign muons are required to be separated by
$\Delta R(\Pgm,\Pgm)=\sqrt{\smash[b]{\Delta\eta^2+\Delta\phi^2}}>2$,
where $\Delta\eta$ is the separation in pseudorapidity and $\Delta\phi$
is the separation in azimuthal angle between the two muons.
If more than one same-sign muon pair
is found in the event, the pair with the largest scalar sum of muon transverse momenta is chosen.

The analysis makes use of reconstructed tracks that fulfill selection criteria 
based on the track fit quality, the number of measurements in the inner pixel and
outer strip silicon detector, and track impact parameters with respect to
the PV~\cite{TRK-11-001}.  
Tracks must have $\pt>1\GeV$ and $|\eta|<2.4$.
The impact parameter in the transverse plane and the longitudinal impact parameter 
are required to be smaller than 1\cm relative to the PV.

Given the search topology, we require each muon to be accompanied by exactly one track
satisfying these criteria within a $\Delta R$ cone of radius 0.5 centred on the muon direction.
We label such muon-track pairs as ``isolated''.

The loose impact parameter requirements on the tracks are designed to suppress background events in which 
a heavy-flavour hadron decays into a muon and several charged particles. 
Although tracks from these decay products will be displaced from the PV, they can still satisfy 
the loose track impact parameter criteria. Such events are rejected by the requirement of exactly one track accompanying the muon.

The track around each muon is identified 
as a one-prong \Pgt\ lepton decay candidate if it fulfils the following selection criteria.
\begin{itemize}
\item{The nearby track is required to have charge opposite to the muon.}
\item{The track must have $\pt > 2.5\GeV$ and $|\eta| < 2.4$.}
\item{The transverse and longitudinal impact parameters of the track are required to 
be smaller than 200\micron and 400\micron relative to the PV, respectively.}
\end{itemize} 

\section{Signal extraction}
\label{Sec:SignalExtraction}
The set of selection requirements outlined in the previous section
defines the signal region.
The number of selected data events, the expected background and signal yields, 
and the signal acceptances after selection in the signal region
are reported in Table~\ref{tab:signal_selection}. 
The expected background and signal yields, along with the signal acceptances, are obtained from simulation.
The signal yields are normalized to the benchmark value of the signal production cross
section times branching fraction 
of 5\unit{pb}. The quoted uncertainties in predictions from simulation include only MC statistical uncertainties.
It should be noted that
no MC simulation is used to evaluate the background in the analysis described below 
as the modelling is based fully on data.
The expected background yields presented in Table~\ref{tab:signal_selection}
show that the final selected sample is dominated by QCD multijet events,
and that the contribution from other background sources is negligible, constituting less than 1\% of all selected events.
Although MC simulation is not directly used to estimate background, the simulated samples
play an important role in the validation of the background modelling
as described in Section~\ref{Sec:Bkgd}.  
The signal acceptances are computed with respect to all possible decays of the four \Pgt\ leptons, and
include a branching fraction factor 

\begin{displaymath}
\frac{1}{2}\mathcal{B}^{2}(\phi_{1} \to \Pgt_\Pgm\Pgt_\text{one-prong})\approx 3.5\%,
\end{displaymath}

where the factor $1/2$ accounts for the selection of same-sign muon pairs, and 
$\mathcal{B}(\phi_{1} \to \Pgt_\Pgm\Pgt_\text{one-prong})$ denotes the branching fraction of the $\phi_{1}\to\Pgt\Pgt$ 
decays 
to the final states characterized by the presence of only two charged particles where at least one of the charged particles 
is a muon. This branching fraction is expressed as

\begin{displaymath}
\mathcal{B}(\phi_{1} \to \Pgt_\Pgm\Pgt_\text{one-prong}) = 
2\mathcal{B}(\Pgt \to\text{one-prong})\mathcal{B}(\Pgt\to\Pgm\PGn\PAGn)-\mathcal{B}^{2}(\Pgt\to\Pgm\PGn\PAGn),
\end{displaymath}

where $\mathcal{B}(\Pgt \to\text{one-prong})$  denotes the total branching fraction of the \Pgt\ decay to 
one charged particle with any number of neutral particles.
The factor of two in the first term
accounts for the two possible charges of the required muonic decay: 
$\Pgt^-\Pgt^+ \to \Pgm^- + \text{one-prong}^+$ and $\Pgt^-\Pgt^+ \to \Pgm^+ + \text{one-prong}^-$.
Subtraction of the term $\mathcal{B}^{2}(\Pgt \to \Pgm\PGn\PAGn)$ avoids double counting 
in the case where the two \Pgt~leptons produced by a given $\phi_1$ both decay to muons.

\begin{table*}[tb]
\begin{center}
\topcaption{
The number of observed events, expected background and signal yields, and
signal acceptances after final selection. 
The computed signal acceptances include
the branching fraction factor 
$\mathcal{B}^{2}(\phi_{1} \to \Pgt_\Pgm\Pgt_\text{one-prong})/2$.
The electroweak background contribution includes the Drell--Yan process, $\PW+\mathrm{jets}$ production, 
and diboson production of \PW\PW, \PW\cPZ, and \cPZ\cPZ.
The numbers of signal events are reported for the benchmark value of the signal production cross section times branching fraction of 5\unit{pb}.
The expected background and signal yields and signal acceptances are obtained from simulation.
The quoted uncertainties in predictions from simulation include only statistical uncertainties related to the size of MC samples.
}
\begin{tabular}{llcc}
\hline 
\\ [-12pt]
\multicolumn{2}{l}{ \multirow{2}{*}{Sample} }  & Signal acceptance & \multirow{2}{*}{Number of events}  \\
  & & ${\cal{A}}(\cPg\cPg\to \PH(125)\to \phi_{1}\phi_{1} \to 4\Pgt)$  & \\ 
  \\ [-12pt]
  \hline 
  \\ [-10pt]
  \multicolumn{2}{l}{Signal}  & & for $\sigmaBR = 5\unit{pb}$ \\
   
& { $m_{\phi_{1}}=4\GeV$ }
 & $(5.38\pm0.23)\times 10^{-4}$ 
 & $53.0\pm2.3$ 
 \\
& { $m_{\phi_{1}}=5\GeV$ }
 & $(4.36\pm0.21)\times 10^{-4}$ 
 & $43.0\pm2.0$ 
 \\
& { $m_{\phi_{1}}=6\GeV$ }
 & $(4.00\pm0.23)\times 10^{-4}$ 
 & $39.5\pm2.0$ 
 \\
& { $m_{\phi_{1}}=7\GeV$ }
 & $(4.04\pm0.20)\times 10^{-4}$ 
 & $39.9\pm2.0$ 
 \\
& { $m_{\phi_{1}}=8\GeV$ }
 & $(3.13\pm0.18)\times 10^{-4}$ 
 & $30.8\pm1.8$ 
 \\
 \\ [-10pt]
\hline 
\\ [-10pt]
\multicolumn{2}{l}{Background}   \\
\\ [-12pt]
& {QCD multijet}  & --- & $820\pm320$  \\
& \ttbar        & --- & $1.2\pm0.2$   \\
& {Electroweak} &  ---  & $5.0\pm4.7$  \\
\\ [-10pt]
\hline 
\\ [-10pt]
\multicolumn{2}{l}{Data} & --- & 873  \\
\\ [-10pt]
\hline
\end{tabular}
\label{tab:signal_selection}
\end{center}
\end{table*}

\begin{figure*}[hbtp]
\begin{center}
\includegraphics[width=0.6\textwidth]{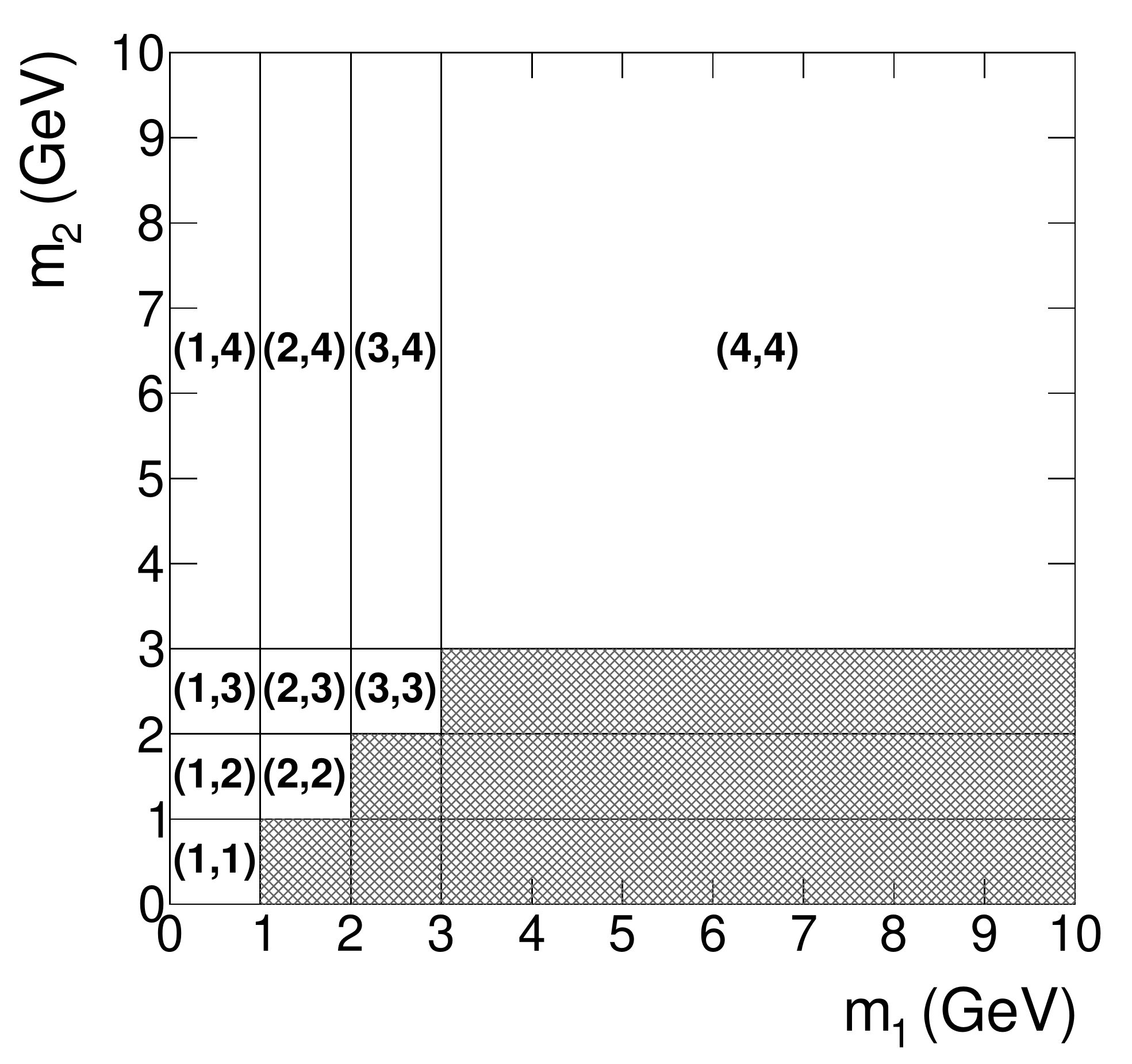}
\caption{
Binning of the two-dimensional ($m_1,m_2$) distribution.
The hatched bins are excluded from the statistical analysis, as detailed in the text.}
\label{fig:binning}
\end{center}
\end{figure*}

The invariant mass of each selected muon and the nearby track is reconstructed.
The two-dimensional distribution of the invariant mass of each selected muon and the nearby track
is used to discriminate between the signal and the QCD multijet background; the signal is extracted by 
means of a fit to this two-dimensional distribution. 
The binning of the two-dimensional ($m_1,m_2$) distributions is illustrated in Fig.~\ref{fig:binning}.
For masses below 3\GeV, bins of 1\GeV width are used for both $m_1$ and $m_2$.
For masses in the range $3 < m_1(m_2) < 10\GeV$, a single bin is used. This choice avoids
poorly populated bins in the two-dimensional ($m_1,m_2$) distributions
in the background control regions used to construct and validate the QCD multijet background model (Section 6).
For each selected event, the ($m_1,m_2$) histogram is filled once if the pair of quantities
($m_1,m_2$) occurs in one of the diagonal bins and twice, once with values ($m_1,m_2$) and a second
time with the swapped values ($m_2,m_1$), for off-diagonal bins. This procedure insures the symmetry of the
two-dimensional ($m_1,m_2$) distribution.
To avoid double counting of events,
the off-diagonal bins $(i,j)$ with $i>j$ are 
excluded from the procedure of the signal extraction (the hatched bins in Fig.~\ref{fig:binning}).
Thus, the number of independent bins is reduced from $4\times 4=16$ to $4\times (4+1)/2=10$.

In order to fit the data in the 10 bins of the two-dimensional
distribution of Fig.~\ref{fig:binning}, a two-component fit is performed using two-dimensional
distributions (``templates'') describing the QCD multijet background and the signal. The normalisations of background and signal
components are free parameters in this fit. The two-dimensional template for the signal is obtained from the simulation using the
generator described in Section~\ref{sec:detector}. The two-dimensional template for the QCD multijet background is extracted from the data as explained in
the next section.

\section{Modelling of the QCD multijet background shape}
\label{Sec:Bkgd}
A simulation study shows that the sample of same-sign 
dimuon events selected as described in Section~\ref{Sec:Selection}, but without
requiring a presence of one-prong \Pgt~candidates and without 
applying the isolation requirement for the muon-track systems, 
is dominated by QCD multiparton production, where 
94\% of all selected events contain \PQb quarks in the final state.
The same-sign muon pairs in these events originate mainly 
in the following cases.
\begin{itemize}
\item{Muonic decay of a bottom hadron in one \PQb quark jet, and 
cascade decay of a bottom hadron into a charmed hadron with subsequent 
muonic decay of a charmed hadron in the other \PQb quark jet.}
\item{Muonic decay of a bottom hadron in one \PQb quark jet,
and decay of a quarkonium state into a pair of muons in the other 
jet.}
\item{Muonic decay of a bottom hadron in one \PQb quark jet,
and muonic decay of a neutral B meson in the other \PQb quark jet.
The same-sign muon pair in this case may appear as a result of $\PBz$--$\PaBz$ oscillations.}   
\end{itemize}

The normalization of the QCD multijet background is not constrained
prior to the extraction of the signal.
The procedure used to model the shape
of the two-dimensional ($m_1,m_2$) distribution
of QCD multijet events in the signal region is described in this section.

Given the symmetry of the two-dimensional ($m_1,m_2$) distribution,  
the modelling of the QCD multijet background shape is derived from the 
two-dimensional probability density function (pdf)

\begin{equation}
f_\text{2D}(m_1,m_2)=C(m_1,m_2)f_\text{1D}(m_1)f_\text{1D}(m_2),
\label{eq:MassSignal}
\end{equation}

where 
\begin{itemize}
\item{$f_\text{2D}(m_1,m_2)$ is the two-dimensional pdf
of the invariant masses of the muon-track systems, $m_1$ and $m_2$,
in the sample of QCD multijet events selected in the signal region;}
\item{$f_\text{1D}(m_i)$ is the one-dimensional pdf of
the invariant mass of the muon-track system in the sample of QCD multijet events selected in the signal region;}
\item{$C(m_1,m_2)$ is a symmetric function of two arguments, $C(m_1,m_2)=C(m_2,m_1)$, reflecting the correlation between $m_1$ and $m_2$.} 
\end{itemize}
A constant correlation function
would indicate the absence of correlation between $m_1$ and $m_2$.
Based on Eq.~(\ref{eq:MassSignal}), the content of bin 
$(i,j)$ of the symmetric normalized two-dimensional distribution $f_\text{2D}(m_1,m_2)$ is 
computed as

\begin{equation}
f_\text{2D}(i,j) = C(i,j) \,f_\text{1D}(i) \, f_\text{1D}(j) ,
\label{eq:QCDshape}
\end{equation}

where
\begin{itemize}
\item $C(i,j)$ is the correlation coefficient in the bin $(i,j)$ of the correlation function $C(m_1,m_2)$;
\item $f_\text{1D}(i)$ is the content of bin $i$ in the normalized one-dimensional distribution $f_\text{1D}(m)$.
\end{itemize}

The modelling of $f_\text{1D}(m)$ and $C(m_1,m_2)$, described in the following, is necessary in order to build the template
$f_\text{2D}(i,j)$.

\subsection{Modelling of \texorpdfstring{$f_\text{1D}(m)$}{f[1D](m)}}

The $f_\text{1D}(m)$ pdf is modelled using a QCD-enriched control data sample disjoint 
from the signal region. Events in the control
sample are required to satisfy all selection criteria, 
except for the isolation of the second muon-track system. The second muon is 
required to be accompanied by either two or three nearby tracks
with $\pt>1\GeV$ and impact parameters smaller than 1\cm relative to the PV
both in the transverse plane and along the beam axis. 
The simulation shows that more than 99\% of events selected in
this control region, hereafter referred to as {\bf{N$_{23}$}}, 
are QCD multijet events.
The modelling of the $f_\text{1D}(m)$ pdf is based on the assumption that the kinematic distributions for the 
first muon-track system are not affected by the isolation requirement imposed on the second, 
and therefore the $f_\text{1D}(m)$ pdf
of the isolated muon-track system is the same in the signal region and the region {\bf{N$_{23}$}}. 

\begin{figure*}[hbtp]
\begin{center}
\includegraphics[width=0.49\textwidth]{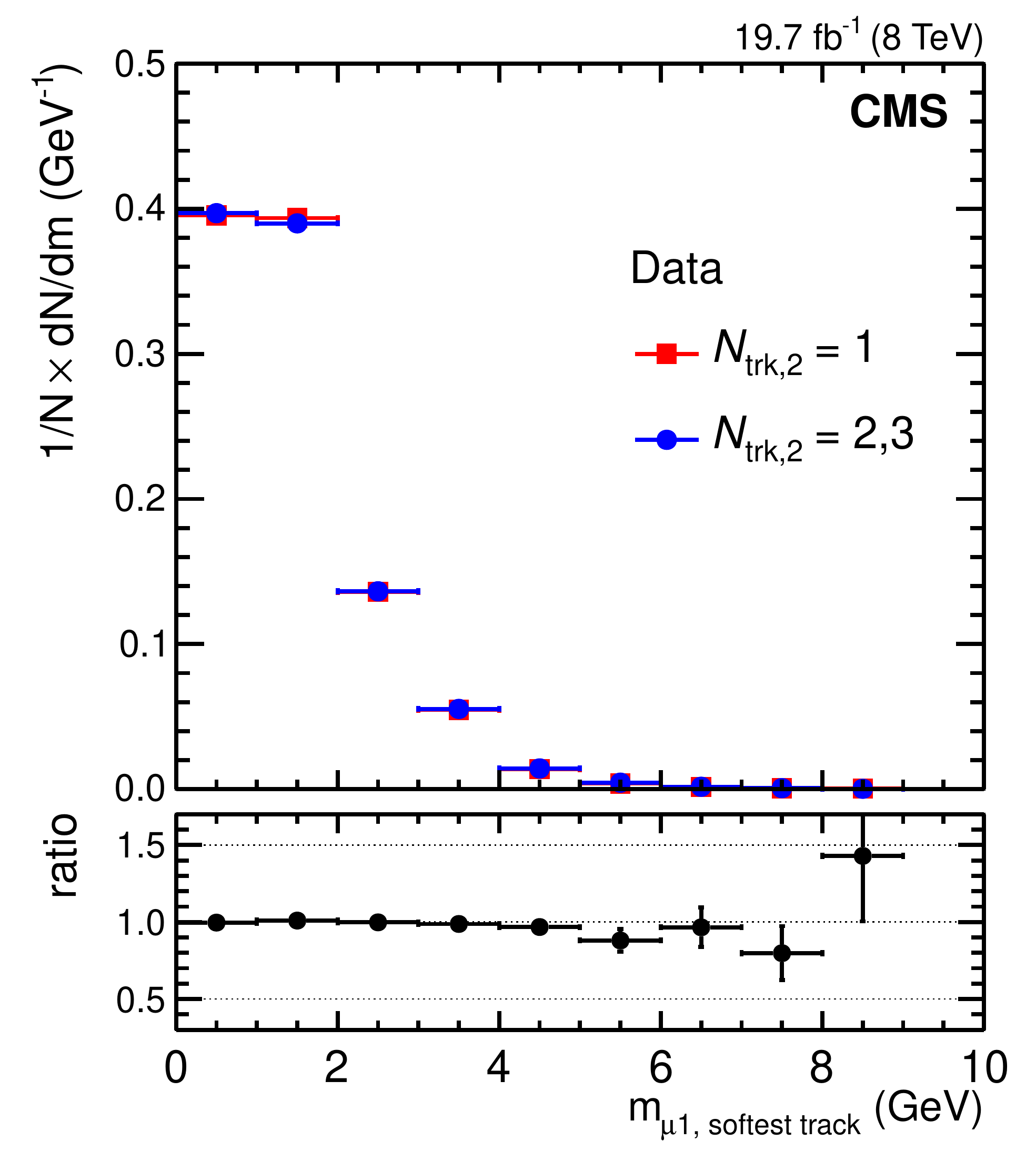} 
\includegraphics[width=0.49\textwidth]{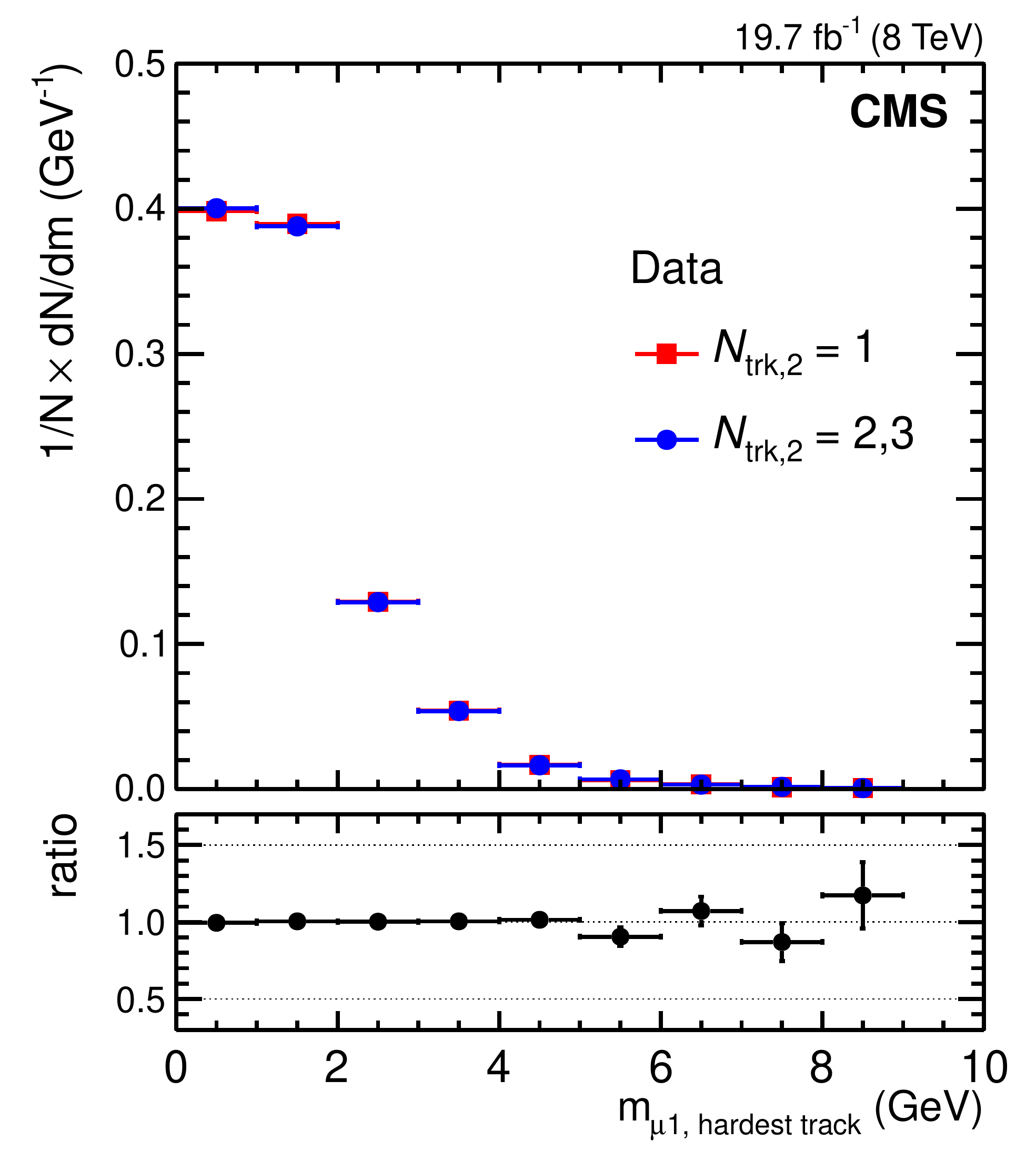}
\vspace{5mm} 
\includegraphics[width=0.49\textwidth]{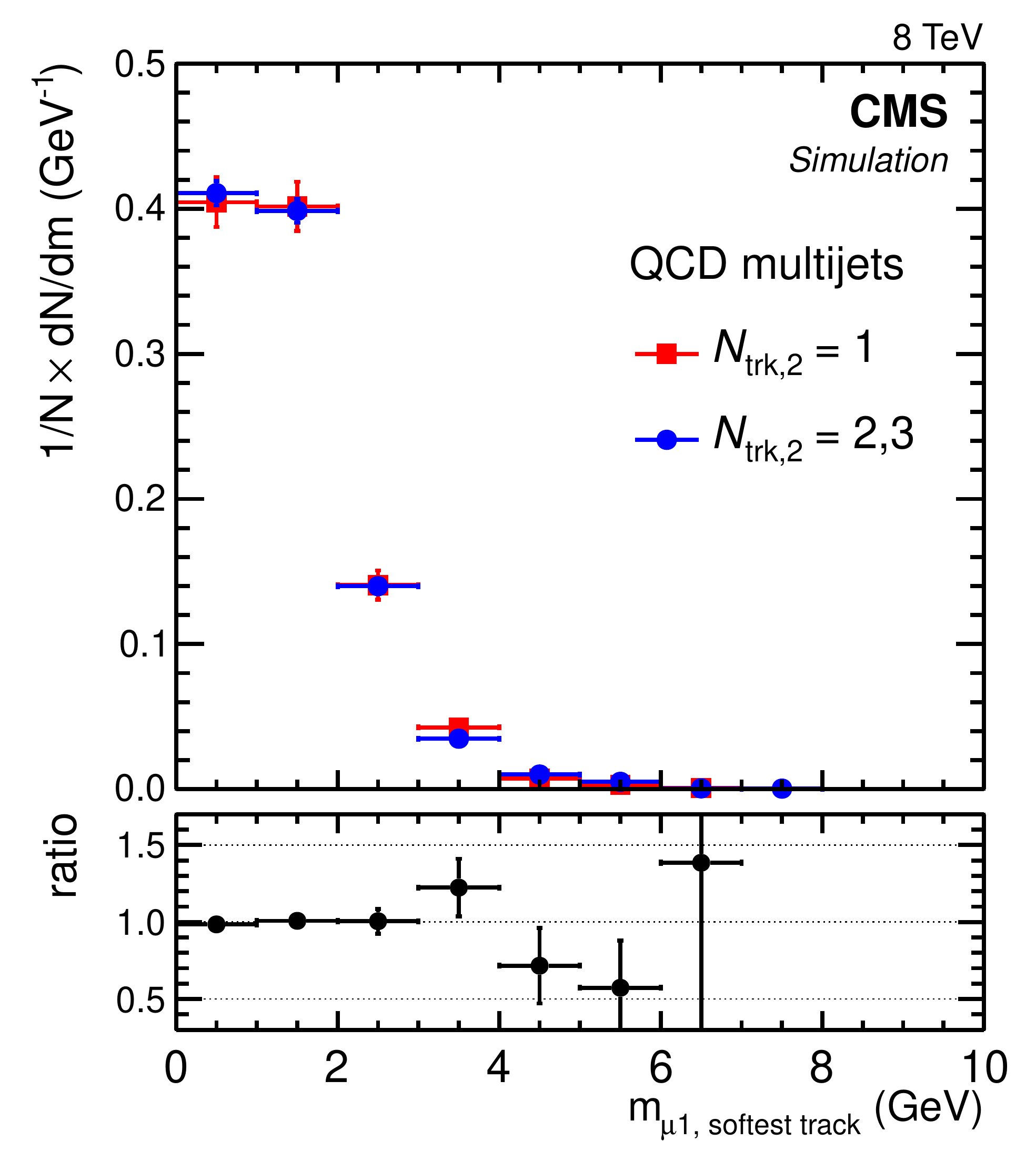} 
\includegraphics[width=0.49\textwidth]{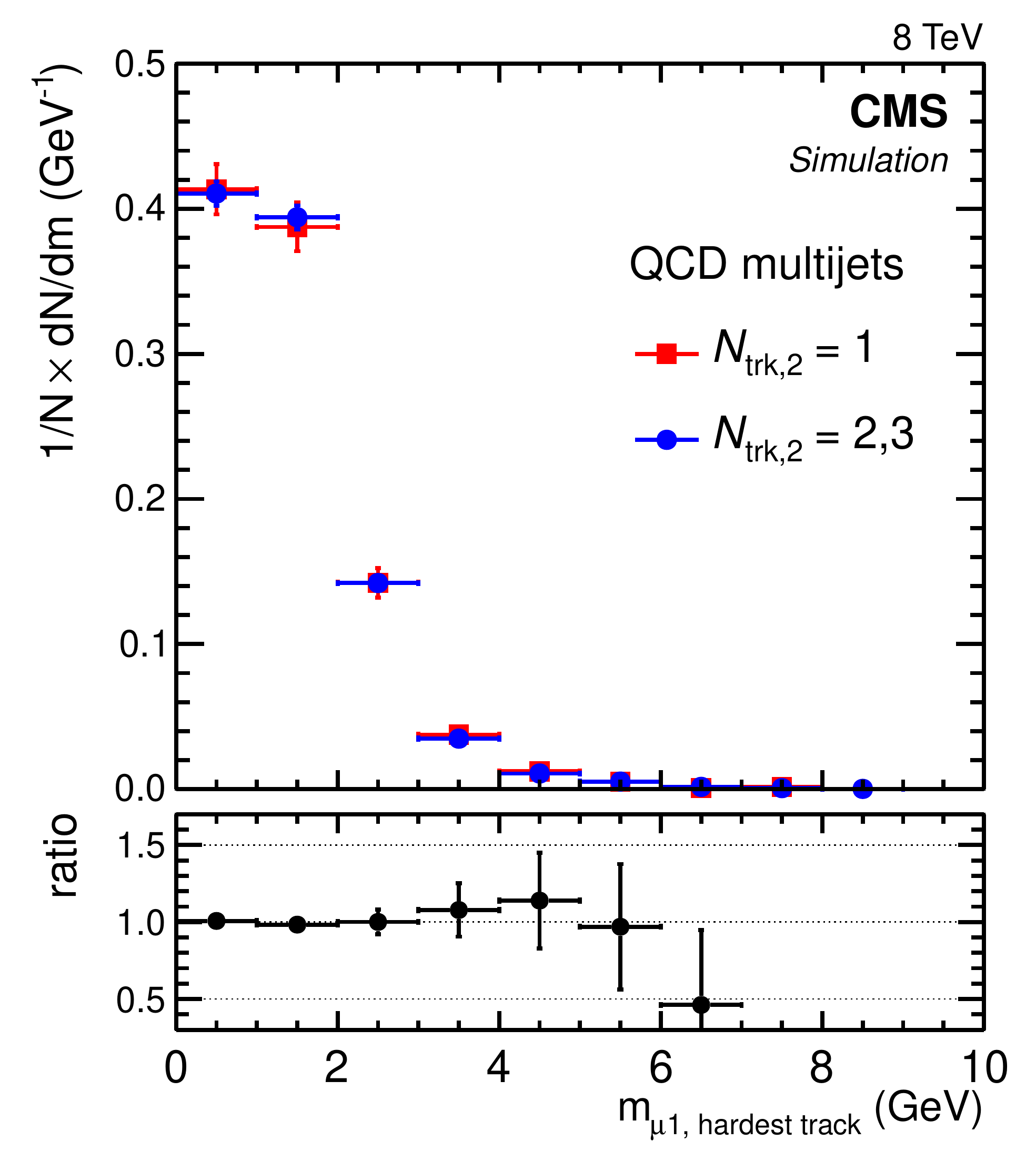} 
\caption{
Normalized invariant mass distributions of the first muon and the softest (left plots) or
hardest (right plots) accompanying track 
for different isolation requirements imposed on the second muon: 
when the second muon has only one accompanying track ($N_\mathrm{trk,2}=1$; squares);
or when the second muon has two or three accompanying tracks ($N_\mathrm{trk,2}=2,3$; circles).
The upper plots show distributions obtained from data. The lower plots show
distributions obtained from the sample of QCD multijet events
generated with \PYTHIA. Lower panels in each plot show the ratio of the
$N_\mathrm{trk,2}=1$ distribution to the $N_\mathrm{trk,2}=2,3$
distribution.}
\label{fig:MassShapesIso}
\end{center}
\end{figure*}

A direct test of this assumption, given the limited size of the simulated 
sample of QCD multijet events, is not conclusive, and a test is therefore performed with
an additional control sample. Events
are selected in this control sample if one of the muons has at least one track passing the one-prong \Pgt\ decay 
candidate criteria within a $\Delta R$~cone of radius 0.5 around the muon direction,
with any number of additional tracks within the same $\Delta R$~cone.
As more than one of these tracks can pass the selection criteria for a one-prong \Pgt\ decay candidate, we investigate two scenarios. In one
scenario, the lowest \pt (``softest'') track passing the one-prong \Pgt\ decay candidate
criteria is used to calculate the muon-track invariant mass, while in the other scenario
the highest \pt (``hardest'') track passing the one-prong \Pgt\ decay candidate criteria is used.
If only one \Pgt\ track is found around 
the first muon, the track is regarded as both ``hardest'' and ``softest''.
For the second muon, two isolation requirements are considered: when the muon is accompanied
by only one track passing the one-prong \Pgt\ decay candidate criteria ($N_\mathrm{trk,2}=1$) 
as in the signal region, or when it is accompanied by two or three tracks ($N_\mathrm{trk,2}=2,3$) 
with $\pt>1\GeV$ and impact parameters smaller than 1\cm relative to the PV
as in the region {\bf{N$_{23}$}}.
The shapes of invariant mass distributions of the first muon and the softest or hardest accompanying track
are then compared for the two different isolation requirements on the second muon, 
$N_\mathrm{trk,2}=1$ and $N_\mathrm{trk,2}=2,3$.
The test is performed both on data and on the simulated sample of
QCD multijet events. The results of this study are illustrated in Fig.~\ref{fig:MassShapesIso}. In all considered cases, the shape of
the invariant mass distribution is compatible within statistical uncertainties between the two cases, 
$N_\mathrm{trk,2}=1$ and $N_\mathrm{trk,2}=2,3$. This observation validates
the assumption that the $f_\text{1D}(m)$ pdf can be determined in the control region {\bf{N$_{23}$}}.

\begin{figure*}[hbtp]
\begin{center} 
\includegraphics[width=0.75\textwidth]{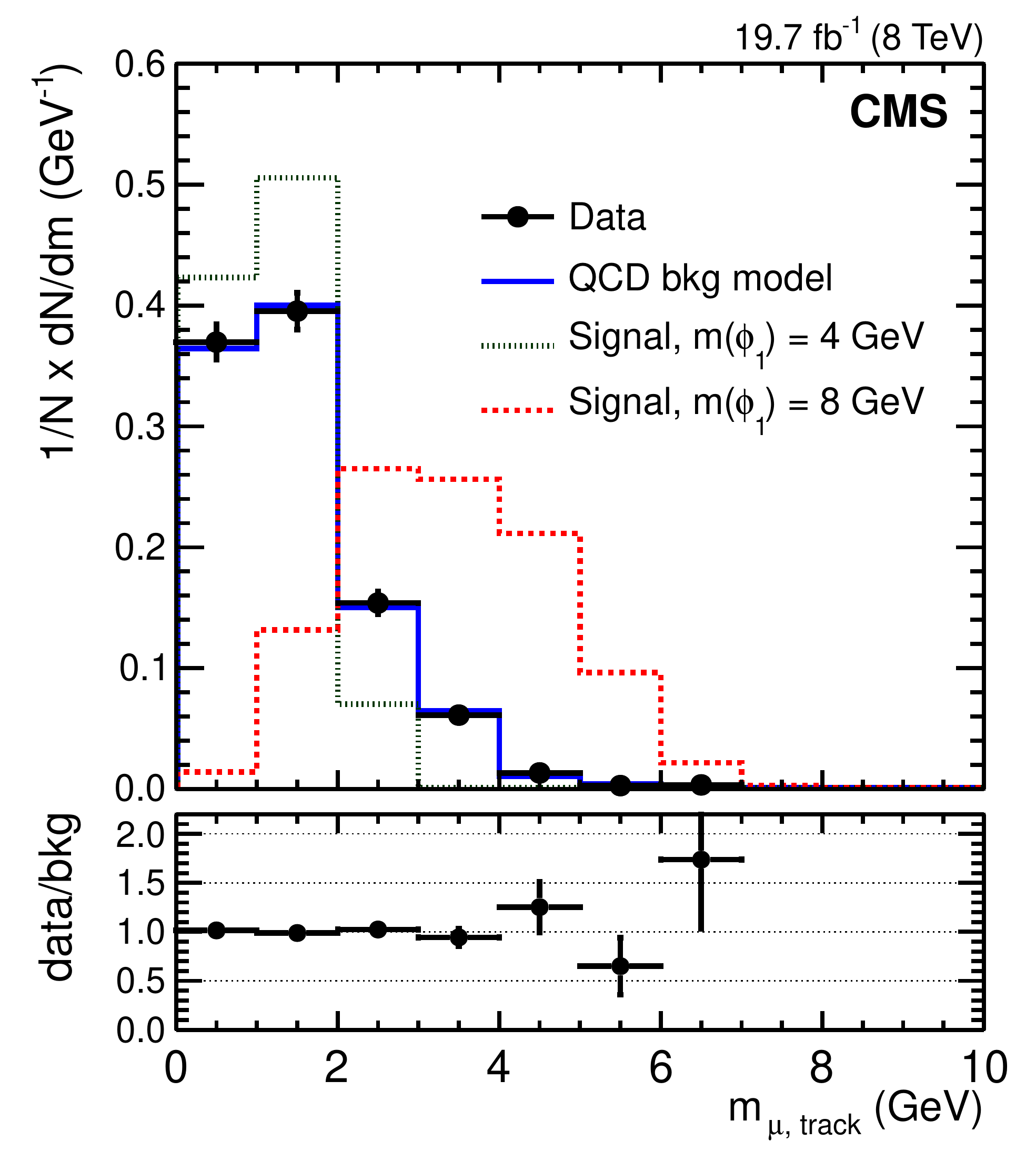}
\end{center} 
\caption{
Normalized invariant mass distribution of the muon-track system for events passing the signal selection.
Data are represented by points. The QCD multijet background model is derived from the
control region {\bf{N$_{23}$}}.
Also shown are the normalized distributions from signal simulations for
two mass hypotheses, $m_{\phi_1}=4\GeV$ (dotted histogram) and 8\GeV (dashed histogram).
Each event contributes two entries to the distribution, 
corresponding to the two muon-track systems passing 
the selection requirements.
The lower panel shows the ratio of the distribution
observed in data to the distribution, describing the
background model.}
\label{fig:MassShapes}
\end{figure*}

Figure~\ref{fig:MassShapes} presents the normalized
invariant mass distribution of the muon-track system for
data selected in the signal region, and for the QCD multijet background model derived from the control region {\bf{N$_{23}$}}.
The data and QCD multijet background distributions are
compared to the signal distribution 
normalized to unity (signal pdf), obtained from simulation, 
for two representative mass hypotheses, $m_{\phi_1}=4$ and 8\GeV.
The invariant mass of the muon-track system is found to have high discrimination power
between the QCD multijet background and signal at $m_{\phi_1}=8\GeV$.
At smaller $m_{\phi_1}$ the signal shape becomes more similar to the background shape,
resulting in a reduction of discrimination power. The normalized
distribution $f_\text{1D}(i)$ with the binning defined in 
Fig.~\ref{fig:binning} is extracted from the background distribution shown in Fig.~\ref{fig:MassShapes}.

\subsection{Modelling of \texorpdfstring{$C(m_1,m_2)$}{C(m[1],m[2])}}

\begin{figure*}[hbtp]
\begin{center}
\includegraphics[width=0.6\textwidth]{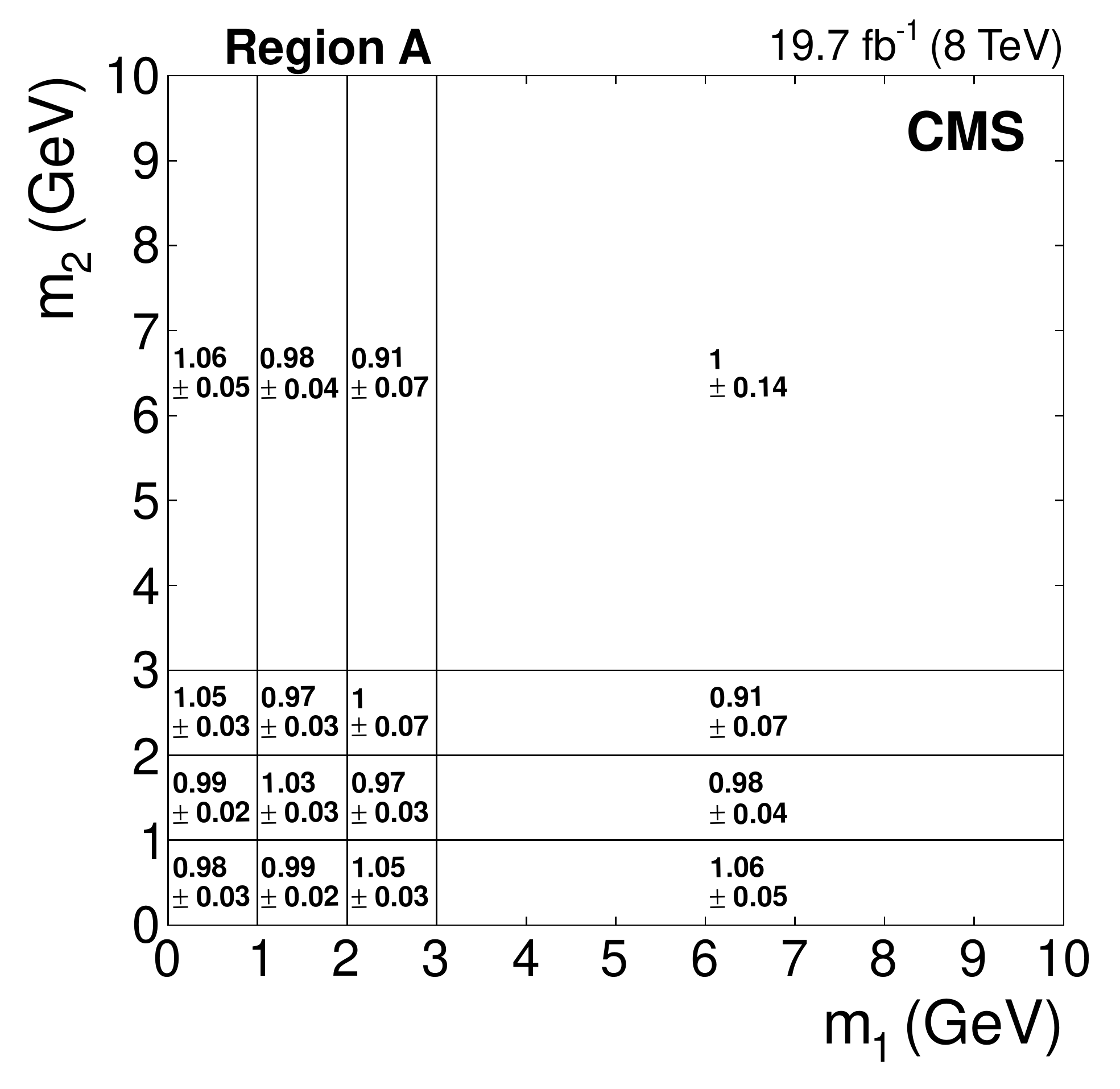}
\caption{
The ($m_1,m_2$) correlation coefficients $C(i,j)$ along with their statistical uncertainties, 
derived from data in the control region {\bf{A}}.}
\label{fig:CorrCoeff}
\end{center}
\end{figure*}

In order to determine the correlation coefficients $C(i,j)$ we define an additional control region {\bf{A}} enriched in 
QCD multijet events. This control region consists of events that contain two same-sign muons
passing the identification and kinematic selection criteria outlined in Section~\ref{Sec:Selection}. 
Each muon is required to have
two or three nearby tracks within a $\Delta R$ cone of radius 0.5 around the muon direction. One and only one 
of these tracks must
satisfy the criteria imposed on one-prong \Pgt\ lepton decay candidates with $\pt>2.5\GeV$. 
The additional tracks must have transverse momentum 
in the range $1 < \pt < 2.5\GeV$.  A total of 9127 data events are selected in this control region.
The MC simulation predicts that the QCD multijet background dominates in region {\bf{A}},
comprising more than 99\% of all selected events. The simulation study also shows
that the overall background-to-signal ratio is enhanced compared to the signal region by a factor of
15 to 20, depending on the mass hypothesis $m_{\phi_1}$. Despite the large increase in the overall background-to-signal ratio,
potential signal contamination in individual bins of the mass distributions
can be nonnegligible. Bin-by-bin signal contamination in region {\bf{A}} is discussed in Section~\ref{Sec:Systematics}.
For each event in control region {\bf{A}}, the pair ($m_1$,$m_2$) of muon-track 
invariant masses is calculated. This pair is used to 
build the symmetrized normalized two-dimensional distribution $f_\text{2D}(i,j)$ defined in 
Fig.~\ref{fig:binning}. Then $C(i,j)$ is obtained according to Eq.~(\ref{eq:QCDshape}) as

\begin{equation} 
C(i,j) = \frac{f_\text{2D}(i,j)}{f_\text{1D}(i) \, f_\text{1D}(j)},
\end{equation}

where $f_\text{1D}(i)$ is the one-dimensional normalized
distribution with two entries per event ($m_1$ and $m_2$) built as for 
Fig.~\ref{fig:MassShapes}.
Correlation coefficients $C(i,j)$ derived from data in region {\bf{A}} are presented in Fig.~\ref{fig:CorrCoeff}.

A direct comparison of $C(i,j)$ between the signal region and region {\bf{A}} 
would be impossible in the simulated sample of QCD multijet events 
because of the very small numbers of events selected in the signal region 
and in region {\bf{A}}. 
In order to assess the difference in $C(i,j)$ between the signal region and 
region {\bf{A}}, a dedicated MC study is performed, making use of a
large exclusive sample generated with \PYTHIA.
The simulation includes only two
leading-order QCD multijet production mechanisms:
the creation of a \bbbar\ quark pair via 
$\Pg\Pg \to \bbbar$ and 
$\Pq\Paq \to \bbbar$.
The detector simulation and event reconstruction are not performed
for this sample, and the comparison of $C(i,j)$ between 
the signal region and region {\bf{A}} is made using generator-level quantities.

\begin{figure*}[hbtp]
\begin{center}
\includegraphics[width=0.6\textwidth]{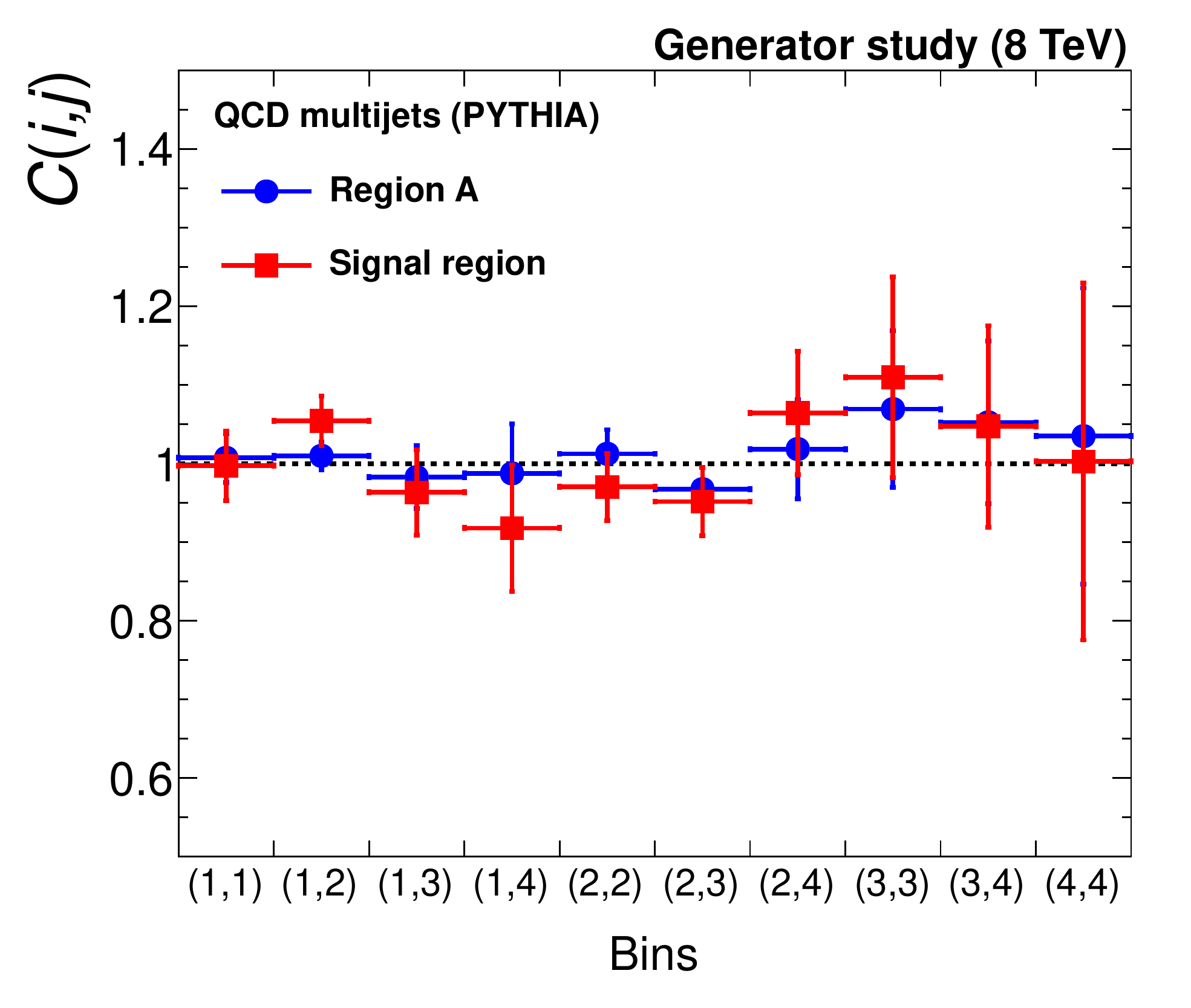}
\caption{
The ($m_1,m_2$) correlation coefficients $C(i,j)$ determined in the control region {\bf{A}} (circles) 
and in the signal region (squares) from the MC study carried out at generator level
with the exclusive MC sample of QCD multijet events resulting from 
$\Pg\Pg(\Pq\Paq) \to \bbbar$ production mechanisms. 
The bin notation follows the definition presented in Fig.~\ref{fig:binning}. 
The vertical bars include both statistical and systematic uncertainties.}
\label{fig:massCorr}
\end{center}
\end{figure*}

These simplifications are validated by performing a set of consistency tests, making use of the available MC sample of
QCD multijet events processed through the full detector simulation and event reconstruction. These tests are performed in 
a control region {\bf{B}}, where each muon is required to have at least one
track passing the one-prong \Pgt\ decay candidate selection criteria, i.e.\ with $\pt>2.5\GeV$ and 
impact parameters smaller than 200\micron and 400\micron in the transverse plane and along beam axis, respectively.  
Along with this requirement each muon is allowed to have one or more tracks within 
a $\Delta R$ cone of radius 0.5 around the muon direction, with $\pt>1\GeV$ and 
impact parameters smaller than 1\cm. Control region {\bf{B}} is characterized by a significantly larger yield of QCD multijet events
compared to the signal region and control region {\bf{A}}, thus making it possible to perform
reliable MC consistency tests and assess the uncertainties in $C(i,j)$.
Two scenarios are investigated: 1) muons are paired with the softest one-prong \Pgt\ decay candidate and
2) muons are paired with the hardest one-prong \Pgt\ decay candidate. If only one one-prong \Pgt\ decay
candidate is found around a muon, it is regarded as both ``softest'' and ``hardest''. 
In both scenarios the correlation coefficients computed using the reconstructed four-momenta of muons and tracks are found to be
compatible with those computed using generator-level four-momenta, within statistical uncertainties. 
Furthermore, the correlation coefficients
computed with the inclusive QCD multijet sample are found to be compatible 
with those computed in the exclusive MC sample including only the
$\Pg\Pg(\Pq\Paq) \to \bbbar$ production mechanisms. This observation validates the use of the generator-level information
and the exclusive \bbbar\  MC sample to compare
$C(i,j)$ between the signal region and control region {\bf{A}}. This comparison is presented in Fig.~\ref{fig:massCorr}.
The uncertainties in $C(i,j)$ represent a quadratic sum of the systematic and MC statistical uncertainties.
The systematic uncertainties are derived from the control region {\bf{B}}. 
They take into account 1) any differences in $C(i,j)$ calculated using
the inclusive QCD multijet sample compared with the exclusive \bbbar\ sample and 2) any differences in $C(i,j)$ 
calculated using full detector simulation and event reconstruction compared with the study using generator-level quantities.
Within their uncertainties the correlation coefficients $C(i,j)$ in the signal region and in region {\bf{A}} are compatible.
We therefore use $C(i,j)$ derived from data in region {\bf{A}} to predict the QCD multijet
background shape in the signal region according to Eq.~(\ref{eq:QCDshape}).

\section{Systematic uncertainties}
\label{Sec:Systematics}
The analysis is affected by various systematic uncertainties, which are classified into two groups.
The first group consists of uncertainties related to the background, while the second group includes uncertainties related to the signal.
The systematic uncertainties are summarized in Table~\ref{tab:systematics}.

\begin{table*}[hb]
\begin{center}
\topcaption{Systematic uncertainties and their effect on the estimates of the
QCD multijet background and signal. The effect of the uncertainties in $C(i,j)$ on
the total background yield is absorbed by the overall background normalization, which is
allowed to vary freely in the fit.}
\label{tab:systematics}
\begin{tabular}{lcccc}
\hline
\\ [-12pt]
\multirow{2}{*}{Source} & \multirow{2}{*}{Value} & Affected & \multirow{2}{*}{Type} & Effect on the  \\
       &       & sample   &      & total yield    \\
\\ [-12pt]
\hline
\\ [-12pt]
Statistical               & \multirow{2}{*}{2--14\%}       & \multirow{2}{*}{bkg.}   & \multirow{2}{*}{bin-by-bin} & \multirow{2}{*}{---}  \\
uncertainties in $C(i,j)$ &               &        &            &     \\
\\ [-10pt]
Extrapolation             & \multirow{2}{*}{2--22\%}       & \multirow{2}{*}{bkg.}   & \multirow{2}{*}{bin-by-bin} & \multirow{2}{*}{---} \\
uncertainties in $C(i,j)$ &               &        &            &    \\
\\ [-10pt] 
Integrated luminosity     & 2.6\%         & signal & norm. & 2.6\% \\
\\ [-10pt] 
Muon ID and trigger       & \multirow{2}{*}{2\% per muon}  & \multirow{2}{*}{signal} & \multirow{2}{*}{norm.} & \multirow{2}{*}{4\%} \\
efficiency                &               &        &       &     \\
\\ [-10pt] 
Track selection and       & \multirow{2}{*}{5\% per track} & \multirow{2}{*}{signal} & \multirow{2}{*}{norm.} & \multirow{2}{*}{10\%} \\
isolation efficiency      &               &        &       &     \\
\\ [-10pt]
MC statistical            & \multirow{2}{*}{7--100\%}      & \multirow{2}{*}{signal} &  \multirow{2}{*}{bin-by-bin}  & \multirow{2}{*}{4--6\%} \\
uncertainties             &               &        &              & \\
\\ [-12pt]
\hline
\\ [-5pt]
\multicolumn{5}{c}{Theory uncertainties in the signal acceptance} \\
\hline
\\ [-10pt]
$\mu_\text{r}$ and $\mu_\text{f}$ variations & 1\%   & signal & norm. & 1\% \\
\\ [-10pt]
PDF              & 1\%   & signal & norm. & 1\% \\
\\ [-10pt]
Effect of \PQb quark loop                    &  \multirow{2}{*}{3\%}  & \multirow{2}{*}{signal} & \multirow{2}{*}{norm.} & \multirow{2}{*}{3\%} \\
contribution to $\cPg\cPg \to \PH(125)$   &       &        &       &     \\
\\ [-12pt]
\hline
\end{tabular}
\end{center}
\end{table*}

\subsection{Uncertainties related to background}

The estimation of the QCD multijet background is based solely on data and is therefore not
affected by imperfections in the simulation of the detector response and 
inaccuracies in the modelling of the muon and track reconstruction.

The shape of the background in the two-dimensional 
($m_1,m_2$) distribution is modelled according
to Eq.~(\ref{eq:QCDshape}). The uncertainty in the two-dimensional shape 
$f_\text{2D}(m_1,m_2)$ is dominated by uncertainties in the correlation coefficients $C(i,j)$ derived
in the QCD multijet background-enriched control region {\bf{A}} as described in Section~\ref{Sec:Bkgd}.
The statistical uncertainties in $C(i,j)$ in region {\bf{A}}
range from 2 to 14\%, as seen in Fig.~\ref{fig:CorrCoeff}.
These uncertainties are accounted for in the signal extraction procedure 
by 10 independent nuisance parameters,
one nuisance parameter per bin in the  ($m_1,m_2$) distribution.
The systematic uncertainties related to the extrapolation of $C(i,j)$ from the control region {\bf{A}}
to the signal region are derived from the dedicated MC study. The correlation coefficients
are found to be compatible between the signal region and the control region {\bf{A}}
within uncertainties ranging from 2 to 22\% (Fig.~\ref{fig:massCorr}). 
These uncertainties are accounted for by 10 additional independent nuisance parameters.

The possible contamination of control region {\bf{A}} by the signal
may bias the estimation of the correlation coefficients and
consequently have an impact on the evaluation of the QCD multijet background.
The effect is estimated with a conservative assumption on the branching fraction
$\mathcal{B}(\PH(125)\to \phi_1\phi_1) \, \mathcal{B}^{2}(\phi_1\to\Pgt\Pgt)$ of 32\%,
which corresponds to the 95\% confidence level (CL) upper limit set by CMS
on the branching fraction of the $\PH(125)$ boson decays to non-standard model 
particles~\cite{CMS_Higgs_properties}, while the cross section for gluon-gluon fusion is set to 
the value predicted in the standard model (19.3\unit{pb}).
Under these assumptions, the contamination of region {\bf{A}} by the signal
is estimated to be less than 2\% for all mass hypotheses $m_{\phi_1}$ and
in all bins of the two-dimensional ($m_1,m_2$) distribution,
with the exception of bin (4,4), where the contamination can reach
12\% for $m_{\phi_1}=8\GeV$.
However, the overall effect on the signal extraction is found to be marginal.
Within this conservative scenario, variations of $C(i,j)$ due to possible
contamination of control region {\bf{A}} by the signal modify
the observed and expected upper limits at 95\% CL on \sigmaBR\
by less than 1\% for all considered values of $m_{\phi_1}$.

\subsection{Uncertainties related to signal}

The following uncertainties in the signal estimate are taken into account, and are summarized in Table~\ref{tab:systematics}.

An uncertainty of 2.6\% is assigned to the integrated luminosity estimate~\cite{CMS-PAS-LUM-13-001}.

The uncertainty in the muon identification and trigger
efficiency is estimated to be 2\% using the tag-and-probe technique applied
to a sample of $\cPZ \to \Pgm\Pgm$ decays.
Because final states with two muons are selected in this analysis,
this uncertainty translates into a 4\% systematic uncertainty in the 
signal acceptance.

The track selection and isolation efficiency is assessed with a study
performed on a sample of \cPZ\ bosons decaying into a pair of \Pgt\ leptons.
In the selected $\cPZ \to \Pgt\Pgt$ events, one \Pgt\ lepton is identified via its muonic decay,
while the other is identified as an isolated track resulting from a one-prong decay.
The track is required to pass the nominal selection criteria
used in the main analysis. From this study the uncertainty in the track selection and 
isolation efficiency is estimated to be 5\%. As the analysis requires each muon
to be accompanied by one track, this uncertainty gives rise to a 10\% systematic uncertainty 
in the signal acceptance.

The muon momentum and track momentum scale uncertainties are smaller than 0.5\% 
and have a negligible effect on the analysis.

The bin-by-bin MC statistical uncertainties in the signal acceptance range from 7 to 100\%.
Their impact on the signal normalization is between 4 and 6\% as 
indicated in Table~\ref{tab:signal_selection}. 
These uncertainties are accounted for in the signal extraction procedure
by 10 nuisance parameters, corresponding to 10 independent bins in the 
($m_1,m_2$) distribution.

Theoretical uncertainties have an impact on the differential kinematic distributions of the
produced $\PH(125)$ boson, in particular its \pt spectrum, thereby affecting signal acceptance.
The uncertainty due to missing higher-order corrections to the gluon-gluon fusion 
process are estimated with the {\sc HqT} program by varying the renormalization ($\mu_\text{r}$) and
factorization ($\mu_\text{f}$) scales. The $\PH(125)$ \pt-dependent $k$ factors are recomputed
according to these variations and applied to the simulated signal samples. The resulting
effect on the signal acceptance is estimated to be of the order of 1\%.  

The {\sc HqT} program is also used to evaluate the effect of the PDF uncertainties. 
The nominal $k$ factors for the $\PH(125)$ boson \pt spectrum
are computed with the MSTW2008nnlo PDF set~\cite{mstw2008}.
Variations of the MSTW2008nnlo PDFs
within their uncertainties change the signal acceptance by about 1\%, whilst using the CTEQ6L1 PDF set 
changes the signal acceptance by about 0.7\%. These variations are covered by the assigned uncertainty of 1\%.

The contribution of \PQb quark loops to the gluon-gluon fusion process depends on the 
NMSSM parameters, in particular $\tan\beta$, the ratio of the vacuum 
expectation values of the two NMSSM Higgs doublets. The corresponding
uncertainty is conservatively estimated by calculating $k$ factors for the $\PH(125)$ boson \pt spectrum
with \POWHEG~\cite{Alioli:2008tz,Nason:2004rx,Frixione:2007vw,Alioli:2010xd},
removing any contribution from the top quark loop and retaining only the contribution from the \PQb quark loop.
The modified $k$ factors applied to the simulated signal samples change the signal acceptance
by approximately 3\% for all mass hypotheses $m_{\phi_1}$.

\section{Results}
\label{Sec:Results}
\newcommand{\pZ}{\ensuremath{\phantom{0}}}
\newcommand{\pZZ}{\ensuremath{\phantom{00}}}
\newcommand{\pF}{\ensuremath{\phantom{00.0}}}
\newcommand{\pFs}{\ensuremath{\phantom{0.0}}}

The signal is extracted with a binned maximum-likelihood fit applied to
the two-dimensional ($m_1,m_2$) distribution in data. For each mass hypothesis of the $\phi_1$
boson, the ($m_1,m_2$) distribution in data is fitted with the QCD multijet background
shape and the $\mathrm{gg\to H(125)}$ signal shape for the $\phi_1$ mass under test.
The contribution to the final selected sample from vector boson fusion and vector boson 
associated production of the $\PH(125)$ boson is suppressed by the selection described in 
Section~\ref{Sec:Selection}, i.e.\ by the requirement $\Delta R(\Pgm,\Pgm) > 2$.
The impact of other backgrounds on the fit is found to be negligible.
The signal shapes are derived from simulation. The background shape 
is evaluated from data, as described in Section~\ref{Sec:Bkgd}.
The systematic uncertainties are accounted for in
the fit via nuisance parameters with log-normal pdfs.

The contribution to the final selected sample from vector boson fusion ($\text{qqH}$) and vector boson
associated production ($\text{VH}$) of the $\PH(125)$ boson is suppressed by the selection described in
Section~\ref{Sec:Selection}, especially by the requirement $\Delta R(\mu,\mu) > 2$.
For the values of the $\PH(125)$ boson production cross sections predicted
in the SM, the expected contribution from the $\text{qqH}$ and
$\text{VH}$ processes to the final selected sample is estimated to be less than 4\%
of total signal yield for all tested $m_{\phi_1}$ hypotheses. The shapes of the
two-dimensional ($m_1,m_2$)
distributions are found to be nearly indistinguishable among the three considered production modes,
making it difficult to extract individual contributions from these processes in a model independent way.
In the following these contributions are neglected, resulting in more
conservative upper limits on \sigmaBR.
Subtraction of the $\text{qqH}$ and $\text{VH}$ contributions assuming the SM cross sections
for the $\PH(125)$ production mechanisms would decrease the upper limits on \sigmaBR\ by less than 4\%
for all tested values of $m_{\phi_1}$.

\begin{figure*}[hbtp]
\begin{center}
\includegraphics[width=0.49\textwidth]{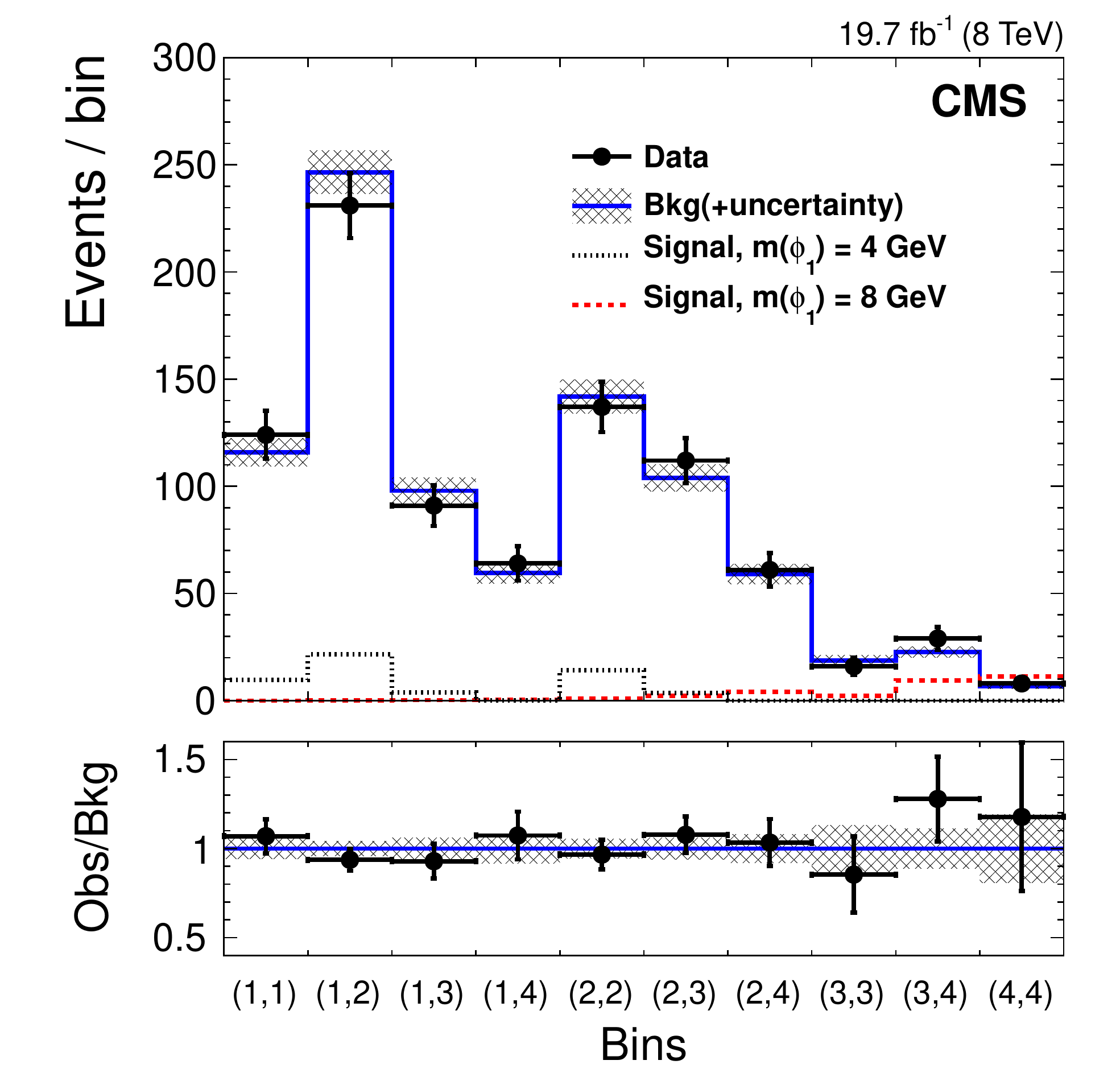}
\includegraphics[width=0.49\textwidth]{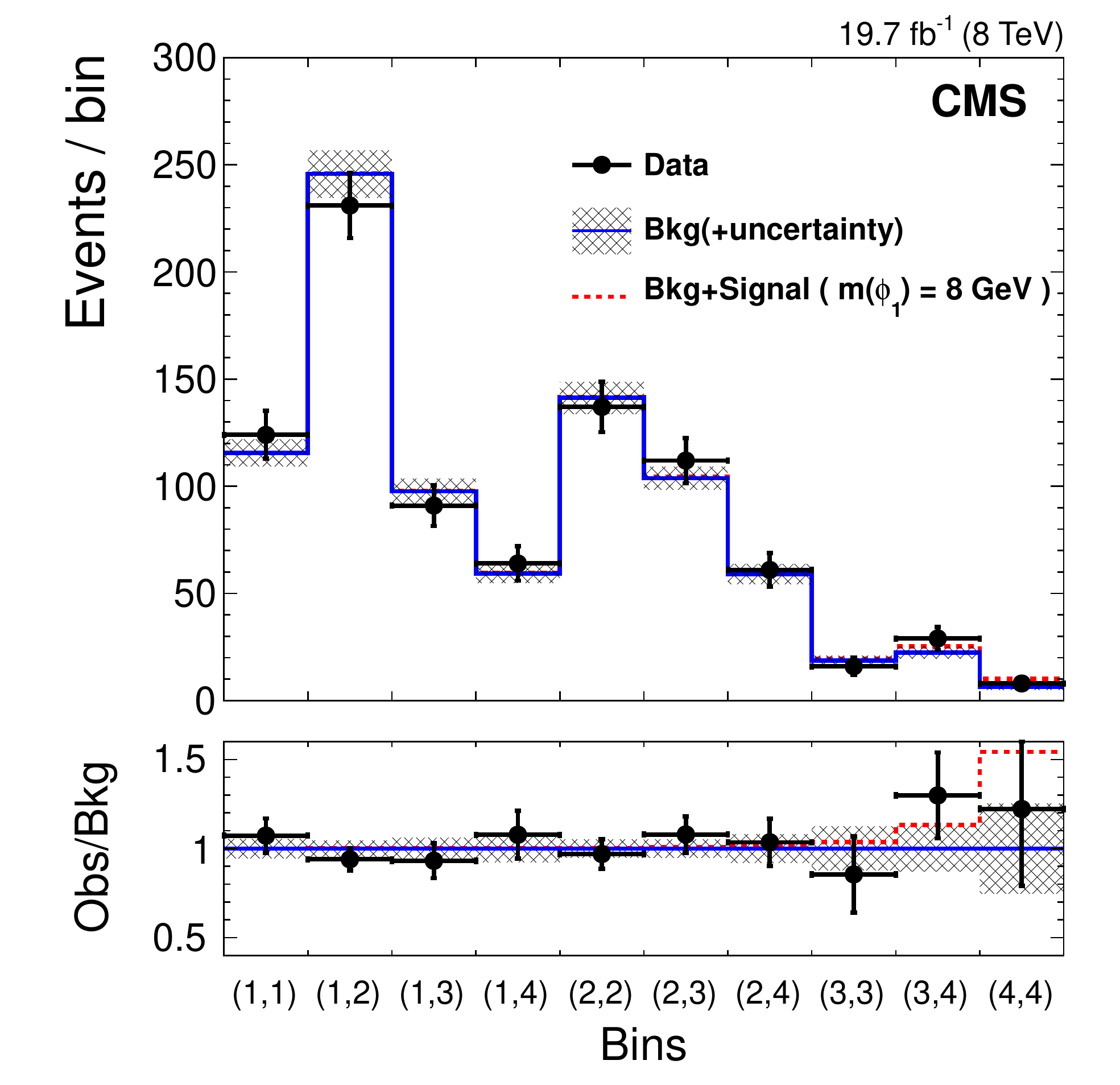}
\caption{
The two-dimensional ($m_1,m_2$) distribution unrolled into a
one-dimensional array of analysis bins.
In the left plot, data (points) are compared with the background prediction (solid histogram)
after applying the
maximum-likelihood fit under the background-only hypothesis and with the signal expectation
for two mass hypotheses, $m_{\phi_{1}}=4$ and 8\GeV (dotted and dashed histograms,
respectively). The signal distributions are obtained from simulation and normalized to a value of the
cross section times branching fraction of 5\unit{pb}.
In the right plot, data (points) are compared with the background prediction (solid histogram) and the background+signal 
prediction for $m_{\phi_{1}}=8\GeV$ (dashed histogram) after applying the maximum-likelihood fit under 
the signal+background hypothesis. The bin notation follows the definition presented in Fig.~\ref{fig:binning}.}
\label{fig:AnalysisBins}
\end{center}
\end{figure*}

First, the data are examined for their consistency with the background-only hypothesis
by means of a fit performed with the normalization of the signal fixed to zero.
Figure~\ref{fig:AnalysisBins} (left) shows the two-dimensional ($m_1,m_2$)
distribution unrolled into a one-dimensional array of analysis bins
after performing the maximum-likelihood fit under the background-only hypothesis. 
The signal distribution, although not used in the fit,
is also included for comparison, for the mass hypotheses $m_{\phi_1}=4$ and 8\GeV.

Table~\ref{tab:YieldBins} presents the number of observed data events, the predicted background yields obtained from a
fit under the background-only hypothesis, and the expected signal yields obtained from simulation, for each unique bin
in the two-dimensional ($m_1,m_2$) distribution. The data are well described by the background-only model.

\begin{table*}[tb]
\begin{center}
\topcaption{The number of observed data events, the predicted background yields, and the expected
signal yields, for different masses of the $\phi_1$ boson in individual bins of the ($m_1,m_2$) distribution.
The background yields and uncertainties are obtained from the maximum-likelihood fit under the background-only
hypothesis. The signal yields are obtained from simulation and normalized to a
signal cross section times branching fraction of 5\unit{pb}. The uncertainties in the signal yields include 
systematic and MC statistical uncertainties.
The bin notation follows the definition presented in Fig.~\ref{fig:binning}.}
\label{tab:YieldBins}
\begin{tabular}{c|ccccccc}
\hline
 \multirow{2}{*}{Bin}     &    \multirow{2}{*}{Data}      &    \multirow{2}{*}{Bkg.}               & 
 \multicolumn{5}{c}{Signal for \sigmaBR~=~5\unit{pb}, $m_{\phi_1} =$ } 
 \\
   &    &           &   4\GeV    &   5\GeV     &  6\GeV     &   7\GeV    &    8\GeV          \\
\hline
(1,1)& $124$    &   $116\pm7\pZ$ &$\pZ 9.7\pm 1.5$&$\pZ 1.9\pm 0.5$&$\pFs {<}0.1 $&$\pZ 0.1\pm 0.1$&$\pF   {<}0.1    $ \\
(1,2)& $231$    &   $247\pm10$   &$   21.6\pm 2.9$&$\pZ 6.8\pm 1.1$&$1.9\pm 0.5$&$\pZ 0.3\pm 0.2$&$\pZ 0.1\pm 0.1$ \\
(1,3)& $\pZ 91$ &$\pZ 98\pm 6\pZ$&$\pZ 3.8\pm 0.8$&$\pZ 4.9\pm 0.9$&$2.4\pm 0.6$&$\pZ 0.9\pm 0.3$&$\pZ 0.2\pm 0.2$ \\
(1,4)& $\pZ 64$ &$\pZ 60\pm 5\pZ$&$\pZ 0.1\pm 0.1$&$\pZ 1.5\pm 0.4$&$1.8\pm 0.5$&$\pZ 0.8\pm 0.3$&$\pZ 0.5\pm 0.2$ \\
(2,2)& $137$    &   $142\pm 8\pZ$&$   14.2\pm 2.0$&$\pZ 8.2\pm 1.3$&$2.8\pm 0.6$&$\pZ 1.5\pm 0.4$&$\pZ 0.8\pm 0.3$ \\
(2,3)& $112$    &   $104\pm 6\pZ$&$\pZ 3.7\pm 0.7$&$   10.4\pm 1.6$&$9.2\pm 1.4$&$\pZ 4.4\pm 0.8$&$\pZ 2.3\pm 0.6$ \\ 
(2,4)& $\pZ 61$ &$\pZ 59\pm 5\pZ$&$\pF  {<}0.1     $&$\pZ 2.6\pm 0.6$&$5.6\pm 1.0$&$\pZ 8.1\pm 1.3$&$\pZ 4.0\pm 0.8$ \\
(3,3)& $\pZ 16$ &$\pZ 19\pm 2\pZ$&$\pF  {<}0.1     $&$\pZ 4.8\pm 0.9$&$4.8\pm 0.9$&$\pZ 3.7\pm 0.7$&$\pZ 2.2\pm 0.5$ \\
(3,4)& $\pZ 29$ &$\pZ 23\pm 3\pZ$&$\pF  {<}0.1     $&$\pZ 1.9\pm 0.5$&$8.0\pm 0.9$&$   11.1\pm 1.5$&$\pZ 9.4\pm 1.4$ \\
(4,4)& $\pZZ 8$ &$\pZZ 7\pm 1\pZ$&$\pF  {<}0.1     $&$\pF    {<}0.1   $&$3.1\pm 0.6$&$\pZ 9.1\pm 1.4$&$   11.2\pm 1.7$ \\
\hline
\end{tabular}
\end{center}
\end{table*}

\begin{figure*}[hbtp]
\begin{center}
\includegraphics[width=0.6\textwidth]{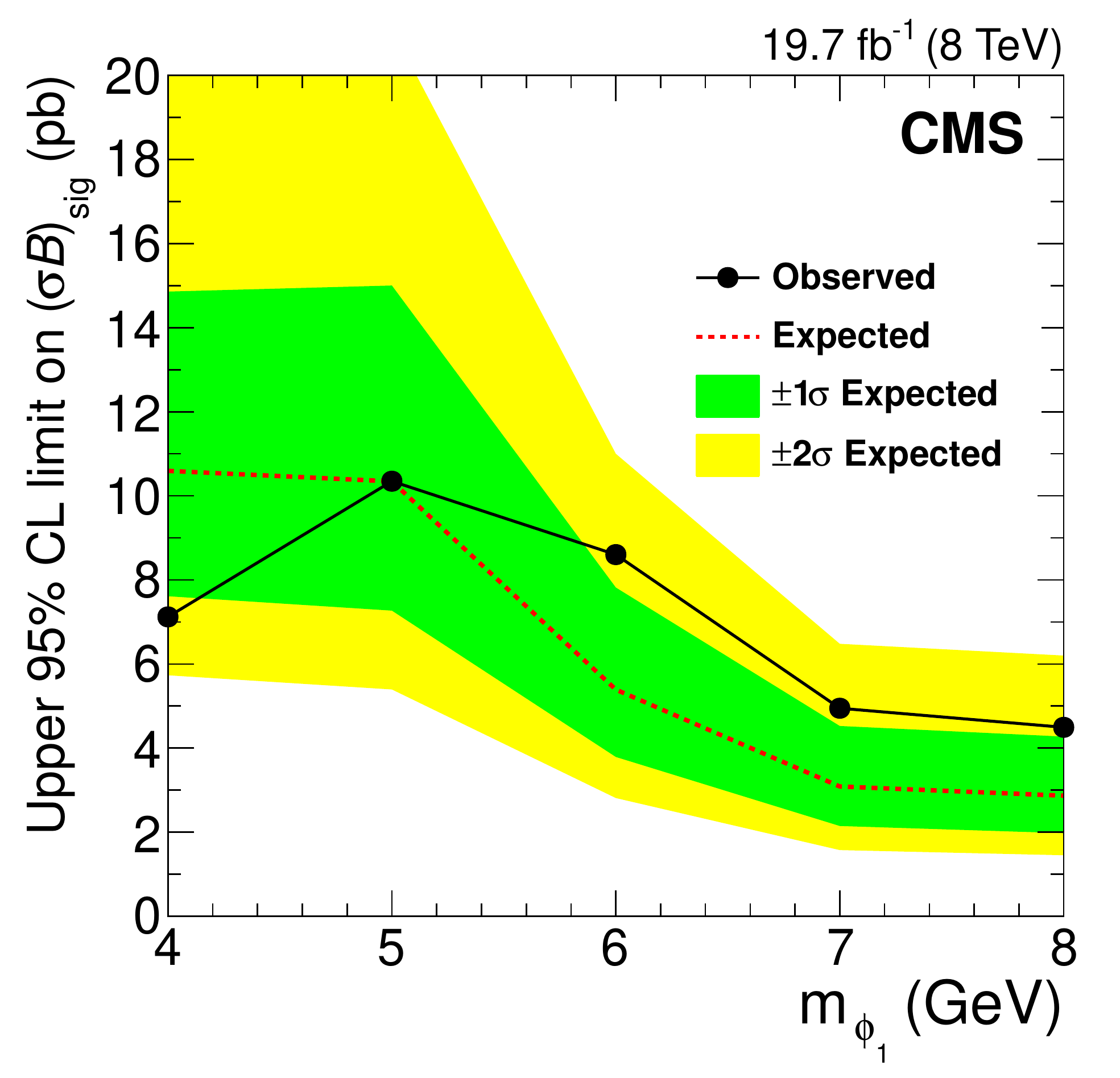}
\caption{
The observed and expected upper limits on \sigmaBR\ in\unit{pb} at 95\% CL, as a function of $m_{\phi_1}$.
The expected limit is obtained under the background-only hypothesis.
The bands show the expected ${\pm}1\sigma$ and ${\pm}2\sigma$ probability intervals around the expected limit.}
\label{fig:limits}
\end{center}
\end{figure*}

The signal cross section times branching fraction is constrained by
performing a fit under the signal+background hypothesis,
where both the background and signal normalisations are allowed to vary freely in the fit.
A representative example of the fit under the signal+background hypothesis at $m_{\phi_1}=8\GeV$
is presented in Fig.~\ref{fig:AnalysisBins} (right). 
No significant deviations from the background expectation 
are observed in data. Only a small excess is found for $6\le m_{\phi_1}\le 8\GeV$,
with a local significance ranging between 1.2$\sigma$ ($m_{\phi_1}=8\GeV$)
and 1.4$\sigma$ ($m_{\phi_1}=6\GeV$).
Results of the analysis are used to set upper limits
on \sigmaBR\ at 95\% CL. The modified frequentist $CL_s$ criterion~\cite{Read:2002hq,Junk:1999kv}, implemented
in the {\sc RooStats} package~\cite{Moneta:2010pm}, is used for the calculation of the exclusion limits.
Figure~\ref{fig:limits} shows the observed upper limit on \sigmaBR\ at 95\% CL,
together with the expected limit obtained under the background-only hypothesis,
for $m_{\phi_1}$ in the range from 4 to 8\GeV.
Exclusion limits are also reported in Table~\ref{tab:limits}.

The observed limit is compatible with the expected limit within two standard deviations
in the entire tested range of the $\phi_1$ boson mass,
$4 \leq m_{\phi_1} \leq 8\GeV$. The observed
limit ranges from 4.5\unit{pb} at $m_{\phi_1}=8\GeV$ to 10.3\unit{pb} at $m_{\phi_1}=5\GeV$.
The expected limit ranges from 2.9\unit{pb} at $m_{\phi_1}=8\GeV$ to 10.6\unit{pb} at $m_{\phi_1}=4\GeV$.

\begin{table*}[tb]
\begin{center}
\topcaption{The observed upper limit on \sigmaBR\
at 95\% CL, together with the expected limit obtained in the background-only hypothesis, as a function of
$m_{\phi_1}$. Also shown are $\pm 1\sigma$ and $\pm 2\sigma$ probability intervals around the expected limit.}
\label{tab:limits}
\begin{tabular}{c|c|ccccc}
\hline
\multirow{2}{*}{$m_{\phi_1} \,\mbox{[{\GeVns}]}$} & \multicolumn{6}{c}{Upper limits on \sigmaBR\ [pb] at 95\% CL} \\
 & observed & $-2\sigma$ & $-1\sigma$ & expected & $+1\sigma$ & $+2\sigma$ \\
\hline
4 & \pZ 7.1 & 5.7 & 7.6 &    10.6 &    14.9 &    20.2 \\
5 &    10.3 & 5.4 & 7.3 &    10.3 &    15.0 &    21.2 \\
6 & \pZ 8.6 & 2.8 & 3.8 & \pZ 5.4 & \pZ 7.8 &    11.0 \\
7 & \pZ 5.0 & 1.6 & 2.2 & \pZ 3.1 & \pZ 4.5 & \pZ 6.5 \\
8 & \pZ 4.5 & 1.5 & 2.0 & \pZ 2.9 & \pZ 4.3 & \pZ 6.2 \\
\hline
\end{tabular}
\end{center}
\end{table*}

The analysis presented here complements the search for $\mathrm{h}/\PH \to \PaO\PaO \to \Pgm\Pgm\Pgt\Pgt$
performed by the ATLAS Collaboration~\cite{Aad:2015oqa}, providing results in 
the $4\Pgt$ channel, which has not been previously explored at the LHC.

\section{Summary}
A search for a very light NMSSM Higgs boson \PaO\ or \PhO, produced in decays of the observed boson with a mass near 125\GeV, $\PH(125)$,
is performed on a \Pp\Pp\ collision data set corresponding to an integrated luminosity of 19.7\fbinv, 
collected at a centre-of-mass energy of 8\TeV.
The analysis searches for the production of an $\PH(125)$ boson via gluon-gluon fusion, and
its decay into a pair of \PaO\ $(\PhO)$ states, each of which decays into a pair of \Pgt\ leptons.
The search covers a mass range of the \PaO\ $(\PhO)$ boson of 4 to 8\GeV. 
No significant excess above background expectations is found
in data, and upper limits at 95\% CL are set on the signal production cross section times branching fraction,

\begin{displaymath}
(\sigma \mathcal{B})_\text{sig} \equiv \sigma (\cPg\cPg \to \PH(125)) \, \mathcal{B} (\PH(125) \to \phi_1 \phi_1) \, \mathcal{B}^{2} (\phi_1 \to \Pgt \Pgt),
\end{displaymath}

where $\phi_1$ is either the \PaO\ or \PhO\ boson.
The observed upper limit at 95\% CL on \sigmaBR\
ranges from 4.5\unit{pb} at $m_{\phi_1}=8\GeV$ to 10.3\unit{pb} at $m_{\phi_1}=5\GeV$.

\section*{Acknowledgements}

\hyphenation{Bundes-ministerium Forschungs-gemeinschaft Forschungs-zentren} We congratulate our colleagues in the CERN accelerator departments for the excellent performance of the LHC and thank the technical and administrative staffs at CERN and at other CMS institutes for their contributions to the success of the CMS effort. In addition, we gratefully acknowledge the computing centres and personnel of the Worldwide LHC Computing Grid for delivering so effectively the computing infrastructure essential to our analyses. Finally, we acknowledge the enduring support for the construction and operation of the LHC and the CMS detector provided by the following funding agencies: the Austrian Federal Ministry of Science, Research and Economy and the Austrian Science Fund; the Belgian Fonds de la Recherche Scientifique, and Fonds voor Wetenschappelijk Onderzoek; the Brazilian Funding Agencies (CNPq, CAPES, FAPERJ, and FAPESP); the Bulgarian Ministry of Education and Science; CERN; the Chinese Academy of Sciences, Ministry of Science and Technology, and National Natural Science Foundation of China; the Colombian Funding Agency (COLCIENCIAS); the Croatian Ministry of Science, Education and Sport, and the Croatian Science Foundation; the Research Promotion Foundation, Cyprus; the Ministry of Education and Research, Estonian Research Council via IUT23-4 and IUT23-6 and European Regional Development Fund, Estonia; the Academy of Finland, Finnish Ministry of Education and Culture, and Helsinki Institute of Physics; the Institut National de Physique Nucl\'eaire et de Physique des Particules~/~CNRS, and Commissariat \`a l'\'Energie Atomique et aux \'Energies Alternatives~/~CEA, France; the Bundesministerium f\"ur Bildung und Forschung, Deutsche Forschungsgemeinschaft, and Helmholtz-Gemeinschaft Deutscher Forschungszentren, Germany; the General Secretariat for Research and Technology, Greece; the National Scientific Research Foundation, and National Innovation Office, Hungary; the Department of Atomic Energy and the Department of Science and Technology, India; the Institute for Studies in Theoretical Physics and Mathematics, Iran; the Science Foundation, Ireland; the Istituto Nazionale di Fisica Nucleare, Italy; the Ministry of Science, ICT and Future Planning, and National Research Foundation (NRF), Republic of Korea; the Lithuanian Academy of Sciences; the Ministry of Education, and University of Malaya (Malaysia); the Mexican Funding Agencies (CINVESTAV, CONACYT, SEP, and UASLP-FAI); the Ministry of Business, Innovation and Employment, New Zealand; the Pakistan Atomic Energy Commission; the Ministry of Science and Higher Education and the National Science Centre, Poland; the Funda\c{c}\~ao para a Ci\^encia e a Tecnologia, Portugal; JINR, Dubna; the Ministry of Education and Science of the Russian Federation, the Federal Agency of Atomic Energy of the Russian Federation, Russian Academy of Sciences, and the Russian Foundation for Basic Research; the Ministry of Education, Science and Technological Development of Serbia; the Secretar\'{\i}a de Estado de Investigaci\'on, Desarrollo e Innovaci\'on and Programa Consolider-Ingenio 2010, Spain; the Swiss Funding Agencies (ETH Board, ETH Zurich, PSI, SNF, UniZH, Canton Zurich, and SER); the Ministry of Science and Technology, Taipei; the Thailand Center of Excellence in Physics, the Institute for the Promotion of Teaching Science and Technology of Thailand, Special Task Force for Activating Research and the National Science and Technology Development Agency of Thailand; the Scientific and Technical Research Council of Turkey, and Turkish Atomic Energy Authority; the National Academy of Sciences of Ukraine, and State Fund for Fundamental Researches, Ukraine; the Science and Technology Facilities Council, UK; the US Department of Energy, and the US National Science Foundation.

Individuals have received support from the Marie-Curie programme and the European Research Council and EPLANET (European Union); the Leventis Foundation; the A. P. Sloan Foundation; the Alexander von Humboldt Foundation; the Belgian Federal Science Policy Office; the Fonds pour la Formation \`a la Recherche dans l'Industrie et dans l'Agriculture (FRIA-Belgium); the Agentschap voor Innovatie door Wetenschap en Technologie (IWT-Belgium); the Ministry of Education, Youth and Sports (MEYS) of the Czech Republic; the Council of Science and Industrial Research, India; the HOMING PLUS programme of the Foundation for Polish Science, cofinanced from European Union, Regional Development Fund; the OPUS programme of the National Science Center (Poland); the Compagnia di San Paolo (Torino); the Consorzio per la Fisica (Trieste); MIUR project 20108T4XTM (Italy); the Thalis and Aristeia programmes cofinanced by EU-ESF and the Greek NSRF; the National Priorities Research Program by Qatar National Research Fund; the Rachadapisek Sompot Fund for Postdoctoral Fellowship, Chulalongkorn University (Thailand); and the Welch Foundation, contract C-1845.

\bibliography{auto_generated}

\cleardoublepage \appendix\section{The CMS Collaboration \label{app:collab}}\begin{sloppypar}\hyphenpenalty=5000\widowpenalty=500\clubpenalty=5000\textbf{Yerevan Physics Institute,  Yerevan,  Armenia}\\*[0pt]
V.~Khachatryan, A.M.~Sirunyan, A.~Tumasyan
\vskip\cmsinstskip
\textbf{Institut f\"{u}r Hochenergiephysik der OeAW,  Wien,  Austria}\\*[0pt]
W.~Adam, E.~Asilar, T.~Bergauer, J.~Brandstetter, E.~Brondolin, M.~Dragicevic, J.~Er\"{o}, M.~Flechl, M.~Friedl, R.~Fr\"{u}hwirth\cmsAuthorMark{1}, V.M.~Ghete, C.~Hartl, N.~H\"{o}rmann, J.~Hrubec, M.~Jeitler\cmsAuthorMark{1}, V.~Kn\"{u}nz, A.~K\"{o}nig, M.~Krammer\cmsAuthorMark{1}, I.~Kr\"{a}tschmer, D.~Liko, T.~Matsushita, I.~Mikulec, D.~Rabady\cmsAuthorMark{2}, B.~Rahbaran, H.~Rohringer, J.~Schieck\cmsAuthorMark{1}, R.~Sch\"{o}fbeck, J.~Strauss, W.~Treberer-Treberspurg, W.~Waltenberger, C.-E.~Wulz\cmsAuthorMark{1}
% ...
\vskip\cmsinstskip
% ...
1:~~Also at Vienna University of Technology, Vienna, Austria\\
2:~~Also at CERN, European Organization for Nuclear Research, Geneva, Switzerland\\
% ...

\end{sloppypar}
\end{document}